\documentclass[10pt,journal,compsoc]{IEEEtran}

\usepackage{comment}
\usepackage{color}
\usepackage{graphicx}
\usepackage{amsmath}
\usepackage{epstopdf}
\usepackage{multicol}
\usepackage{multirow}
\usepackage{multicol}

\begin{document}

\title{Impact and Productivity of PhD Graduates\\of Computer Science/Engineering\\ Departments of Hellenic Universities}

\author{
   Dimitrios Katsaros\ \ \ %$^{1*}$\thanks{*Corresponding author}
   Yannis Manolopoulos%$^{2}$\\
   %\small $^1$Department of Electrical Engineering \& Yale Institute for Network Science, Yale University, USA\\
   %\small $^2$Department of Informatics, Aristotle University of Thessaloniki, Greece\\
   %\small d.katsaros@yale.edu, manolopo@csd.auth.gr
}

\IEEEtitleabstractindextext{
\begin{abstract}
This article presents an anatomy of PhD programmes in Hellenic universities' departments of computer science/engineering from the perspective of research 
productivity and impact. The study aims at showing the dynamics of research conducted in computer science/engineering departments, and after recognizing 
weaknesses, to motivate the stakeholders to take actions that will improve competition and excellence. Beneficiaries of this investigation are the following 
entities: a) the departments themselves can assess their performance relative to that of other departments and then set strategic goals and design procedures 
to achieve them, b) supervisors can assess the part of their research conducted with PhDs and set their own goals, c) former PhDs who can identify their 
relative success, and finally d) prospective PhD students who can consider the efficacy of departments and supervisors in conducting high-impact research 
as one more significant factor in designing the doctoral studies they will follow.
\end{abstract}}

%\markboth{NOT FOR DISTRIBUTION. ONLY FOR INTERNAL CIRCULATION}{Katsaros \& Manolopoulos: PhDs in CSD/ECE departments}

\maketitle

\IEEEpeerreviewmaketitle

\newcommand{\minitab}[2][l]{\begin{tabular}{#1}#2\end{tabular}}

\section{Introduction}
\label{sec-intro}

Nowadays the availability of rich bibliometric data in online databases such as Google Scholar, Elsevier Scopus, Thomson Reuters WoS allows for the data-centric
study of the performance of various entities participating and shaping the research landscape. In an increasing fashion, decision makers related to promotions,
funding, strategic orientation are asking for the exploitation of these data concerning `individuals' (ranging from scientific articles to PhD students, post 
doctorals, faculty members), as well as `collections of individuals' (ranging from journals, to universities, institutions, companies) to backup 
their decisions. There is really huge literature on this topic of data-centric (bibliometric) evaluations, and it is not within the scope of the present article
to survey them (even in a brief manner); instead it directs the interested reader to begin his/her investigation from a couple of recent 
books~\cite{Todeschini-Handbook16,Vitanov-book-science-dynamics}. The use of such quantitative measures in research evaluation is under constant debate; for a 
discussion on the benefits and dangers of research evaluation using bibliometric methods, the reader is directed to~\cite{Gingras-book-abuses} 
and~\cite{Manolopoulos-DAMDID17}.

Departing from all previous national and international studies we investigate in this article the performance of both individuals and organizations from the
perspective of PhD studies. In particular, we compare in terms of productivity, impact and a proxy of both productivity and impact (with the 
$h$-index~\cite{Hirsch-PNAS05}), the lifetime performance of PhDs of Hellenic universities' departments of computer science/engineering, and based on that 
analysis, we evaluate the success of their supervisors and of the respective departments. The salient hypothesis in our study is that the future success of a 
PhD has its origins (at a significant degree) in his/her PhD studies. This hypothesis does not precludes the positive assessment of PhDs who never followed a 
research career after their graduation, because their impact still continues to `grow` and their work can still attract interest, in case it was influential 
and innovative. On the other hand, this hypothesis does not preoccupies the positive assessment of PhDs who followed/follow research careers, because their 
current quantifiable performance might show strong evidence of moderate or even low performance.

The motivation of this study stems from authors' personal interest and their belief that several stakeholders can benefit from this investigation. In particular,
the departments can assess their relative performance and then set strategic goals and design procedures to improve themselves and promote competition and 
excellence. Supervisors can assess the part of their research conducted jointly with PhDs and set their own goals for self-improvement and excellence. Former 
PhDs can identify their relative success and draw conclusions to use in their current jobs, and finally prospective PhD students can consider the efficacy of 
departments and supervisors in conducting high-impact research as one more significant factor in designing the path in their doctoral studies.

The rest of this article is organized as follows: 
Section~\ref{sec-depts-key} describes which departments are evaluated here; Section~\ref{sec-dept-aggr-stats} presents some gross statistics and establishes
the soundness of the present investigation; Section~\ref{sec-supervisor-performance} shows the results which concern the performance of the supervisors, and 
Section~\ref{sec-phds-performance} gives the details of PhDs' performance; Section~\ref{sec-depts-rankings} which is the heart of the present article, 
presents the ranking of the departments; Section~\ref{sec-data-collection} describes our data collection methodology and its challenges; 
Section~\ref{sec-related-work} surveys some related work on evaluations of Hellenic universities' departments, and finally Section~\ref{sec-conclusions} 
concludes this article.

\section{The evaluated departments}
\label{sec-depts-key}

The task of data collection that would feed our study turned out to be quite tough; the challenges are described in Section~\ref{sec-data-collection}. We ended 
up having available data for $15$~university departments, which are shown in Table~\ref{tab-depts-key} sorted according to their key. In the rest of the article, 
this key will be used as the name of the respective department.

\begin{table}[!hbt]
\center
\begin{tabular}{||@{}l|@{}l@{}|@{}c@{}||}\hline\hline
{\bf Key}                  & {\bf Department's name}                                                & {\bf Location}\\\hline
{\scriptsize AEGEAN\_ICSD} & {\tiny Department of Information \& Communication Systems Engineering} & {\scriptsize Samos}\\\hline
{\scriptsize AUEB\_DI}     & {\tiny Department of Informatics}                                      & {\scriptsize Athens}\\\hline
{\scriptsize AUTH\_DI}     & {\tiny Department of Informatics}                                      & {\scriptsize Thessaloniki}\\\hline
{\scriptsize AUTH\_ECE}    & {\tiny Department of Electrical \& Computer Engineering}               & {\scriptsize Thessaloniki}\\\hline
{\scriptsize CRETE\_CSD}   & {\tiny Computer Science Department}                                    & {\scriptsize Heraklion}\\\hline
{\scriptsize CRETE\_ECE}   & {\tiny Department of Electrical \& Computer Engineering}               & {\scriptsize Chania}\\\hline
{\scriptsize IOAN\_CSE}    & {\tiny Department of Computer Science \& Engineering}                  & {\scriptsize Ioannina}\\\hline
{\scriptsize IONIO\_DI}    & {\tiny Department of Informatics}                                      & {\scriptsize Corfu}\\\hline
{\scriptsize NTUA\_ECE}    & {\tiny Department of Electrical \& Computer Engineering}               & {\scriptsize Athens}\\\hline
{\scriptsize PATRAS\_ECE}  & {\tiny Department of Electrical \& Computer Engineering}               & {\scriptsize Patras}\\\hline
{\scriptsize PELOP\_DI}    & {\tiny Department of Informatics and Telecommunications}               & {\scriptsize Tripoli}\\\hline
{\scriptsize UA\_DI}       & {\tiny Department of Informatics and Telecommunications}               & {\scriptsize Athens}\\\hline
{\scriptsize UNIPI\_DI}    & {\tiny Department of Informatics}                                      & {\scriptsize Piraeus}\\\hline
{\scriptsize UTH\_DIB}     & {\tiny Department of Computer Science and Biomedical Informatics}      & {\scriptsize Lamia}\\\hline
{\scriptsize UTH\_ECE}     & {\tiny Department of Electrical \& Computer Engineering}               & {\scriptsize Volos}\\\hline\hline
\end{tabular}
\vspace*{.05\baselineskip}
\caption{Departments studied and their respective key used to represent the department in the plots.}
\label{tab-depts-key}
\end{table}

\begin{figure}[!hbt]
\begin{center}
\includegraphics[scale=.45]{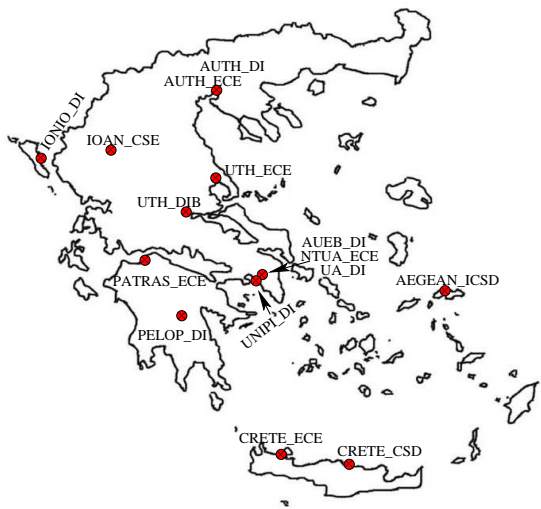}
\end{center}
\vspace*{-\baselineskip}
\caption{Geographical distribution of evaluated departments.}
\label{fig-geo-distribution-of-depts}
\end{figure}

Eight of them are departments which belong to the faculty of sciences in their university (we will call them collectively the `science departments'), 
and the rest belong to the faculty of engineering of their university (we will call them the `engineering departments'). Six out of fourteen departments 
reside in the three major Greek cities, namely Athens, Thessaloniki and Piraeus, and we will call them `the central departments', whereas the rest -- called 
`the peripherals' -- reside in various Greek cities, both in the mainland and in islands. Figure~\ref{fig-geo-distribution-of-depts} shows the geographical 
location of the evaluated departments.

Table~\ref{tab-time-span-of-study} shows the time span covered by {\it our data} for each department. This interval does not necessarily coincides with the 
interval that the respective department has been awarding PhDs. The table also shows the total number of PhDs awarded during that interval.

\begin{table}[!hbt]
\center
\begin{tabular}{||l|c|c||}\hline\hline
{\bf Department} & {\bf Time span of graduations} & {\bf {\#}PhDs awarded}\\\hline
AEGEAN\_ICSD     & 1997--2016                     & $31$  \\\hline
AUEB\_DI         & 1997--2017                     & $77$  \\\hline
AUTH\_DI         & 1997--2016                     & $131$ \\\hline
AUTH\_ECE        & 2005--2016                     & $173$ \\\hline
CRETE\_CSD       & 1992--2016                     & $75$  \\\hline
CRETE\_ECE       & 1992--2017                     & $43$  \\\hline
IOAN\_CSE        & 2005--2017                     & $43$  \\\hline
IONIO\_DI        & 2011--2016                     & $11$  \\\hline
NTUA\_ECE        & 1993--2016                     & $1036$\\\hline
PATRAS\_ECE      & 1984--2017                     & $341$ \\\hline
PELOP\_DI        & 2006--2017                     & $23$  \\\hline
UA\_DI           & 2005--2017                     & $208$ \\\hline
UNIPI\_DI        & 1991--2017                     & $88$ \\\hline
UTH\_DIB         & 2013--2017                     & $6$   \\\hline
UTH\_ECE         & 2006--2017                     & $42$  \\\hline\hline
                 & {\bf TOTAL}                    & {\bf 2328}\\\hline\hline
\end{tabular}
\vspace*{.05\baselineskip}
\caption{Time span of graduations for each department, and number of PhDs graduates covered by our data.}
\label{tab-time-span-of-study}
\end{table}

In other words, we used the productivity and impact data of~$2328$ PhDs awarded from~$15$ departments to carry out evaluation involved persons (PhDs and PhDs 
supervisors) and departments. Our two measures of efficiency are {\it productivity} and {\it impact}. When the term `productivity' is used to describe the output 
of a department or a supervisor, then it refers to the number of awarded PhDs, but when it is used to describe the output of a PhD, then it means the number of 
articles authored by this PhD. We measured the impact of both persons and departments in terms of the number of {\it citations}. Should we had more rich data, 
we could quantify the impact in terms of the prestige of jobs acquired after graduation, in terms of the number of patents granted, in terms of average salaries, 
and so on. However, in this study, we follow the standard bibliometric model by means of citation counting.

We will start presenting our results in the next section by showing how the productivity of each department collectively varies with time, and we will argue 
about the validity of our methodology to compare PhD programmes that have started in different time instances.

\section{A macroscopic view of productivity}
\label{sec-dept-aggr-stats}

\begin{figure}[!hbt]
\begin{center}
\includegraphics[scale=.725]{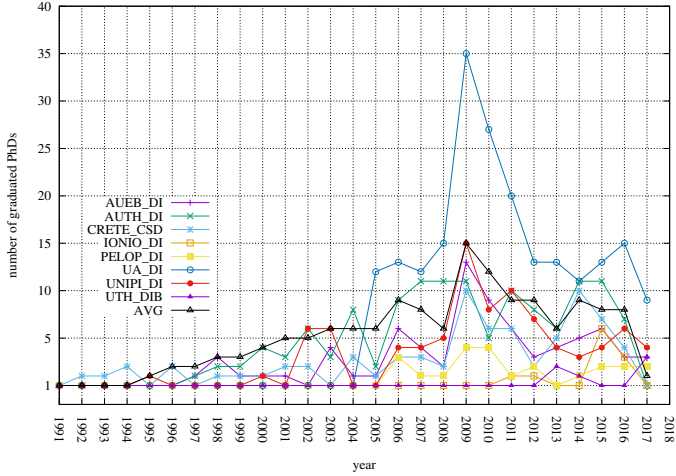}
\end{center}
\vspace*{-\baselineskip}
\caption{Number of PhD graduates from science departments. The average (AVG) is calculated across all departments.}
\label{fig-phds-per-dept-SCI}
\end{figure}

\begin{figure}[!hbt]
\begin{center}
\includegraphics[scale=.725]{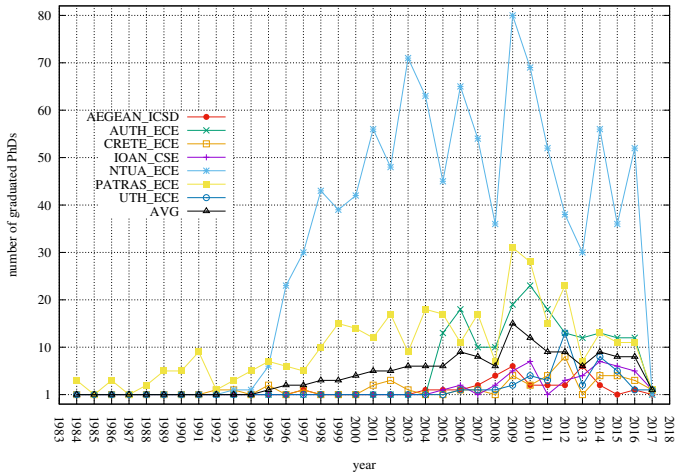}
\end{center}
\vspace*{-\baselineskip}
\caption{Number of PhD graduates from engineering departments. The average (AVG) is calculated across all departments.}
\label{fig-phds-per-dept-ENG}
\end{figure}

Figures~\ref{fig-phds-per-dept-SCI} and~\ref{fig-phds-per-dept-ENG} depict the distribution of awarded PhDs per year and per department for the science and 
engineering departments, respectively. For comparison purposes, we include in each plot the average number of PhDs over {\it all} departments, i.e., both 
science and engineering. The numbers in the plots are absolute numbers without any normalization, e.g., per department's faculty size, because the size and 
composition of faculty members set varies considerably with time, even during the course of a PhD thesis development. In other words, these plots depict the 
`productivity' of the departments in terms of PhD awardees. 

\begin{figure}[!hbt]
\begin{center}
\includegraphics[scale=.65]{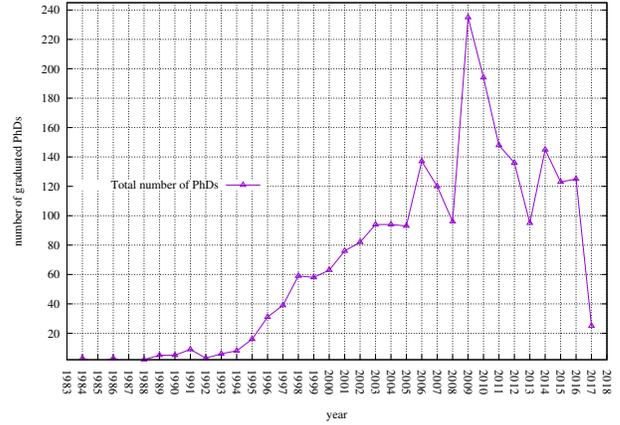}
\end{center}
\vspace*{-\baselineskip}
\caption{Total number of PhD graduates. The data are incomplete for year~$2017$.}
\label{fig-phds-total}
\end{figure}

Obviously, the engineering departments award more PhDs, because they are older and thus they have a) more faculty members, and b) more 'scientifically mature' 
faculty members. The first PhD graduates from science departments appear in~$1991$, compared to~$1984$ from engineering departments. The generic trend is that 
the number of PhDs increases with time. We speculate that this is partly due to social reasons (e.g., increased social recognition), and also due to the 
availability of more funding by the Greek state. This trend was quit steep during the years~$1995-2009$ (see Figure~\ref{fig-phds-total}), and in general 
during the decade of~$2000$, but it declines during the years of recession. We can see that the number of PhDs drops during the years~$2014-2015$, compared to 
the respective number during the period~$2009-2010$. Assuming that a PhD takes~$4$ years to complete, this means that: a) less people are pursuing PhD studies 
during the economic recession, and/or b) departments have less faculty members, and/or c) faculty members are not willing to supervise many PhD students, 
and/or d) faculty members have received less funding, in particular from Greek state agencies.

A couple of science departments, namely AUTH\_DI and UA\_DI are constantly very close or above the average, and the three major engineering departments, namely
NTUA\_ECE, PATRAS\_ECE and AUTH\_ECE are clearly above the average. Looking at the growth pattern, we can observe that almost all (except from the very small
UTH\_DIB) science departments follow the same growth pattern, but only the two major engineering departments (NTUA\_ECE and PATRAS\_ECE) follow their own
pattern.

The average values (depicted in Figure~\ref{fig-phds-per-dept-SCI}-\ref{fig-phds-per-dept-ENG}) show more clearly than Figure~\ref{fig-phds-total}
the gradual decline (mainly after~$2012$) of awarded PhDs, as consequence of the economic recession.

Next, we investigate an issue that comprises the stepping stone for the ranking results presented in Sections~\ref{sec-supervisor-performance}, 
\ref{sec-phds-performance} and~\ref{sec-depts-rankings}. We will investigate the correlation between ranking by impact and ranking by graduation year for the 
PhDs per department. If there is a {\it strong positive} correlation, then any ranking scheme is practically worthless, and we cannot recognize any excellence 
and difference in quality among competitors, because the evolution is governed by the principle of the {\it first-mover's advantage}. On the other hand, if the 
correlation is close to~$0$ or at most ~$0.5$, then this article's results on rankings are sound. In general, this question is fundamental in network science 
research, and represents the battle among {\it the rich gets richer}~\cite{Barabasi-Science99} and {\it the fittest survives}~\cite{Barabasi-Science00} 
principles of evolution.

We use the Kendall's $\tau$ rank correlation coefficient~\cite{Kendall-Biometrika38} to demonstrate how a specific ranking is correlated to another ranking. 
Here, the Kendall's tau coefficient considers the list produced by sorting the PhDs according to their graduation year (the least recent PhDs first), and the 
list produced by sorting the PhDs according to the total number of citations their articles have accumulated. Note that both lists are of the same size i.e., 
$n$. Then, the $\tau$ value can be computed as:
\begin{equation}
\tau = \frac{n_c-n_d}{n(n-1)/2}
\end{equation}
where $n_c$ is the number of concordant pairs, and $n_d$ is the number of discordant pairs. The denominator is the total number of pairs of~$n$ items in the 
lists. For each pair of items in the list, we determine if the relative rankings between the two lists match. For pair of nodes ($i,j$), if node $i$ is ranked 
above (or below) node~$j$ in both lists, then the pair is called {\it concordant}. Otherwise, it is called {\it discordant}. Clearly, $-1 \leq \tau \leq 1$. 
If $\tau=1$, then the two rankings are in perfect agreement; if~$\tau=-1$, then one ranking is the complete reverse of the other. In the ranking list based 
on the graduation year, there are a lot of ties. Therefore, we used as tie breaking criterion the lexicographic ordering of the PhDs' surnames which is a neutral 
criterion. Should we have used the non-increasing number of citations, the correlation coefficient would be slightly higher; if we had used the non-decreasing 
number of citations, then the correlation coefficient would be slightly lower. 

Table~\ref{tab-ranking-correlation} presents the correlation of rankings based on graduation year and impact for the PhDs in each department. We observe that in 
most of the cases (eleven out of fifteen, the correlation coefficient is less than or equal to~$0.5$; this implies that the correlation among the rankings in not 
strong (in a positive way). There are a several cases where the coefficient is very close to~$0$ (approaching it either from left or right) meaning that the two 
rankings are completely uncorrelated. We have even encountered two cases (NTUA\_ECE and UTH\_ECE) where the coefficient is negative, which affirms that recent 
PhDs are (slightly) more impactful, on the 
average.\footnote{For the case of NTUA\_ECE, we need to mention that the magnitude of this negative value might be due to fact that~$70$\% ($30$ out of~$43$) 
of NTUA's unavailable Scopus profiles (cf. Table~\ref{tab-percentages-missing-phds}) is for NTUA\_ECE PhDs who have graduated on or before~$2001$. However, this 
is not the case for UTH\_ECE, where all PhDs are very recent (after~$2006$) and our data are complete, i.e., there are no missing Scopus profiles.}
The only cases, where the correlation can be considered strongly positive are for IONIO\_DI and PEL\_DI which however have not awarded many PhDs and thus the 
statistical sample is not significant. The correlation is very low, i.e., $0.08$ if we consider it across all departments. Therefore, we establish that 
{\it graduation year and impact are not strongly correlated} to each other in a positive or negative way, which is an intuitive result whatsoever.

\begin{table}[!hbt]
\center
\begin{tabular}{||l|c||}\hline\hline
{\bf Department} & {\bf Kendal $\tau$}\\\hline
AEGEAN\_ICSD     &  $0.27$\\\hline
AUEB\_DI         &  $0.17$\\\hline
AUTH\_DI         &  $0.37$\\\hline
AUTH\_ECE        &  $0.34$\\\hline
CRETE\_CSD       &  $0.40$\\\hline
CRETE\_ECE       &  $0.50$\\\hline
IOAN\_CSE        &  $0.58$\\\hline
IONIO\_DI        &  $0.63$\\\hline
NTUA\_ECE        & $-0.04$\\\hline
PATRAS\_ECE      &  $0.03$\\\hline
PELOP\_DI        &  $0.71$\\\hline
UA\_DI           &  $0.39$\\\hline
UNIPI\_DI        &  $0.11$\\\hline
UTH\_DIB         &  $0.54$\\\hline
UTH\_ECE         & $-0.03$\\\hline\hline
ALL\_DEPTS       &  $0.08$\\\hline\hline
\end{tabular}
\vspace*{.05\baselineskip}
\caption{Correlation between graduation year and impact of PhDs per department. The results do not show a strong correlation among them.}
\label{tab-ranking-correlation}
\end{table}

Having established that the graduation year is not really significant to impact, we will refrain from performing any time-based normalization in the results 
shown in the next sections. The next section presents results about the productivity- and impact-based evaluation of supervisors.

\section{Performance of supervisors}
\label{sec-supervisor-performance}

We start our discussion by presenting some plain statistics regarding the number of distinct supervisors, and the average number of PhDs supervised by each one 
of them in all departments. The first issue we need to consider is whether the study should present the data in a(ny) normalized way. For instance, we could show
the average number of PhD graduates of a particular supervisor, where the average would be over the time period starting from the year of first graduate's 
graduation to the year of the last graduate's graduation. We explained in Section~\ref{sec-dept-aggr-stats} why we do not adopt this approach. Moreover, there 
is a second reason for this; such an approach would bias the results towards supervisors who are `active' for very short periods of time. It would be fairer 
to normalize over the whole period of the supervisor's appointment to the specific department; however, this information is not available. Thus, for the
supervisors we choose to present raw `productivity' performance results, but normalized over the number of graduated PhDs for the impact performance results.

Table~\ref{tab-supervisors-giving-phds} depicts the number of faculty members that have supervised at least one PhD graduate in each department. We cannot 
translate these numbers into percentages of the total number of faculty members in each department, because we are lacking the respective data. Besides, the 
number of faculty members in each department varies with time. Comparing the departments AUTH\_DI, AUEB\_DI and CRETE\_CSD with similar number of supervisors, 
we observe that the former department' supervisors are twice as productive as the other two departments.

\begin{table}[!hbt]
\center
\begin{tabular}{||@{}l@{}|@{}c|@{}c||}\hline\hline
{\bf Department} & {\bf {\#}distinct supervisors} & {\bf avg {\#}PhDs per supervisor}\\\hline
AEGEAN\_ICSD     & $8$                            & $3.87$\\\hline
AUEB\_DI         & $25$                           & $3.08$\\\hline
AUTH\_DI         & $22$                           & $5.95$\\\hline
AUTH\_ECE        & $42$                           & $4.11$\\\hline
CRETE\_CSD       & $25$                           & $3$\\\hline
CRETE\_ECE       & $20$                           & $2.15$\\\hline
IOAN\_CSE        & $17$                           & $2.52$\\\hline
IONIO\_DI        & $6$                            & $1.83$\\\hline
NTUA\_ECE        & $89$                           & $11.64$\\\hline
PATRAS\_ECE      & $52$                           & $6.55$\\\hline
PELOP\_DI        & $9$                            & $2.55$\\\hline
UA\_DI           & $41$                           & $5.07$\\\hline
UNIPI\_DI        & $18$                           & $4.88$\\\hline
UTH\_DIB         & $5$                            & $1.2$\\\hline
UTH\_ECE         & $12$                           & $3.5$\\\hline\hline
{\bf TOTAL}      & {\bf 391}                      & \\\hline\hline
\end{tabular}
\vspace*{.05\baselineskip}
\caption{Number of faculty members per department that have supervised at least one PhD graduate.}
\label{tab-supervisors-giving-phds}
\end{table}

Concerning the top-productive supervisors, in Table~\ref{tab-supervisors-productivity-per-dept} we present the top-$3$ for each department, whereas in 
Table~\ref{tab-top-10-productive-supervisors} we present the top-$10$ across all departments. There are cases where the same person-supervisor has supervised 
PhDs in different departments; this is the case for instance of prof.\ Courcoubetis who has supervised PhDs graduates in CRETE\_CSD and AUEB\_DI. We do not 
aggregate these numbers, and ask for the understanding of our esteemed colleagues whose statistics might be slightly affected by this decision.

\begin{table}[!hbt]
\center
\begin{tabular}{||@{}l|@{}c@{}|@{}c@{}|@{}c@{}||}\hline\hline
{\bf Department}           & {\bf top-1}                                                                   & {\bf top-2}                          & {\bf top-3}\\\hline
{\scriptsize AEGEAN\_ICSD} & {\scriptsize Gritzalis, S. ($13$)}                                            & {\scriptsize Kormentzas ($4$)}       & {\scriptsize Kambourakis ($3$)}\\
                           &                                                                               &                                      & {\scriptsize Loukis ($3$)}\\\hline
{\scriptsize AUEB\_DI}     & {\scriptsize Vazirgiannis ($10$)}                                             & {\scriptsize Gritzalis, D. ($9$)}    & {\scriptsize Giannakoudakis ($7$)}\\\hline
{\scriptsize AUTH\_DI}     & {\scriptsize Manolopoulos ($18$)}                                             & {\scriptsize Pitas ($15$)}           & {\scriptsize Vlachavas ($14$)}\\\hline
{\scriptsize AUTH\_ECE}    & {\scriptsize Mitkas ($11$)}                                                   & {\scriptsize Tsiboukis ($10$)}       & {\scriptsize Bakirtzis ($9$)}\\
                           & {\scriptsize Pavlidou ($11$)}                                                 &                                      &              \\
                           & {\scriptsize Strintzis ($11$)}                                                &                                      &              \\\hline
{\scriptsize CRETE\_CSD}   & {\scriptsize Markatos ($8$)}                                                  & {\scriptsize Plexousakis ($7$)}      & {\scriptsize Stylianou ($6$)}\\\hline
{\scriptsize CRETE\_ECE}   & {\scriptsize Zervakis ($7$)}                                                  & {\scriptsize Papaefstathiou ($4$)}   & {\scriptsize Dollas ($3$)}\\
                           &                                                                               &                                      & {\scriptsize Kalaitzakis ($3$)}\\
                           &                                                                               &                                      & {\scriptsize Stavrakakis ($3$)}\\\hline
{\scriptsize IOAN\_CSE}    & {\scriptsize Lykas ($6$)}                                                     & {\scriptsize Nikou ($5$)}            & {\scriptsize Fudos ($4$)}\\
                           &                                                                               &                                      & {\scriptsize Nikolopoulos ($4$)}\\\hline
{\scriptsize IONIO\_DI}    & {\scriptsize Vlamos ($5$)}                                                    & {\scriptsize Andronikos ($2$)}       & \minitab[@{}c@{}]{{\scriptsize Chrissikopoulos ($1$)}\\{\scriptsize Pateli ($1$)}\\{\scriptsize Sioutas ($1$)}\\{\scriptsize Stefanidakis ($1$)}}\\\hline
{\scriptsize NTUA\_ECE}    & {\scriptsize Uzunoglou ($53$)}                                                & {\scriptsize Stasinopoulos ($47$)}   & {\scriptsize Protonotarios($43$)}\\\hline
{\scriptsize PATRAS\_ECE}  & {\scriptsize Goutis ($24$)}                                                   & {\scriptsize Papadopoulos ($23$)}    & {\scriptsize Fakotakis ($19$)}\\\hline
{\scriptsize PELOP\_DI}    & {\scriptsize Simos ($7$)}                                                     & {\scriptsize Vassilakis ($5$)}       & {\scriptsize Vlachos ($3$)}\\\hline
{\scriptsize UA\_DI}       & {\scriptsize Theodoridis, S. ($23$)}                                          & {\scriptsize Merakos ($18$)}         & {\scriptsize Sivridis ($13$)}\\\hline
{\scriptsize UNIPI\_DI}    & \minitab[c]{{\scriptsize Alexandris ($14$)}\\{\scriptsize Douligeris ($14$)}} & {\scriptsize Virvou ($9$)}           & \minitab[@{}c@{}]{{\scriptsize Foundas ($7$)}\\{\scriptsize Tsihrintzis ($7$)}}\\\hline
{\scriptsize UTH\_DIB}     & {\scriptsize Bagos ($2$)}                                                     & \minitab[@{}c@{}]{{\scriptsize Aletras ($1$)}\\{\scriptsize Anagnostopoulos ($1$)}\\{\scriptsize Maglogiannis ($1$)}\\{\scriptsize Plagianakos ($1$)}} & --- \\\hline
{\scriptsize UTH\_ECE}     & {\scriptsize Tassiulas ($14$)}                                                & {\scriptsize Stamoulis, G.I. ($10$)} & {\scriptsize Houstis, E. ($4$)}\\\hline\hline
\end{tabular}
\vspace*{.05\baselineskip}
\caption{Top-$3$ productive supervisors per department. The number in parenthesis is the number of PhDs graduates supervised by the respective person.}
\label{tab-supervisors-productivity-per-dept}
\end{table}

Table~\ref{tab-top-10-productive-supervisors} shows the top-$10$ productive supervisors across all departments. It is expected, and at the same time frustrating 
for the rest of the departments, to see that only supervisors coming from NTUA\_ECE --- which however is the biggest, oldest department --- populate this list.

\begin{table}[!hbt]
\center
\begin{tabular}{||c|c|c||}\hline\hline
{\bf {\#}PhDs} & {\bf supervisor} & {\bf Department}\\\hline
$53$           & Uzunoglou        & NTUA\_ECE\\\hline
$47$           & Stasinopoulos    & NTUA\_ECE\\\hline
$43$           & Protonotarios    & NTUA\_ECE\\\hline
$36$           & Kollias          & NTUA\_ECE\\\hline
$32$           & Sykas            & NTUA\_ECE\\\hline
$29$           & Koutsouris       & NTUA\_ECE\\\hline
$28$           & Venieris         & NTUA\_ECE\\\hline
$27$           & Konstantinou     & NTUA\_ECE\\\hline
$26$           & Theologou        & NTUA\_ECE\\\hline
$25$           & Varvarigou       & NTUA\_ECE\\\hline\hline
\end{tabular}
\vspace*{.05\baselineskip}
\caption{Top-$10$ productive supervisors across all departments.}
\label{tab-top-10-productive-supervisors}
\end{table}

\begin{figure}[!hbt]
\begin{center}
\includegraphics[scale=.3155]{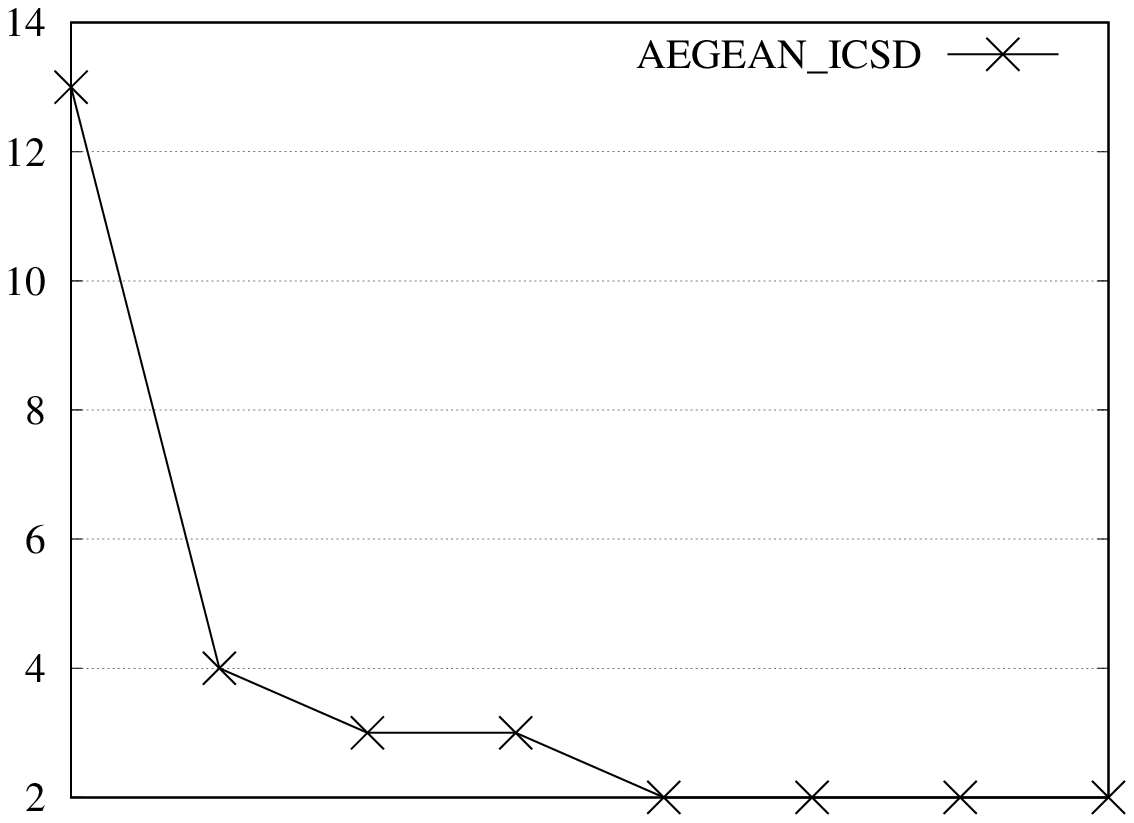}
\includegraphics[scale=.3155]{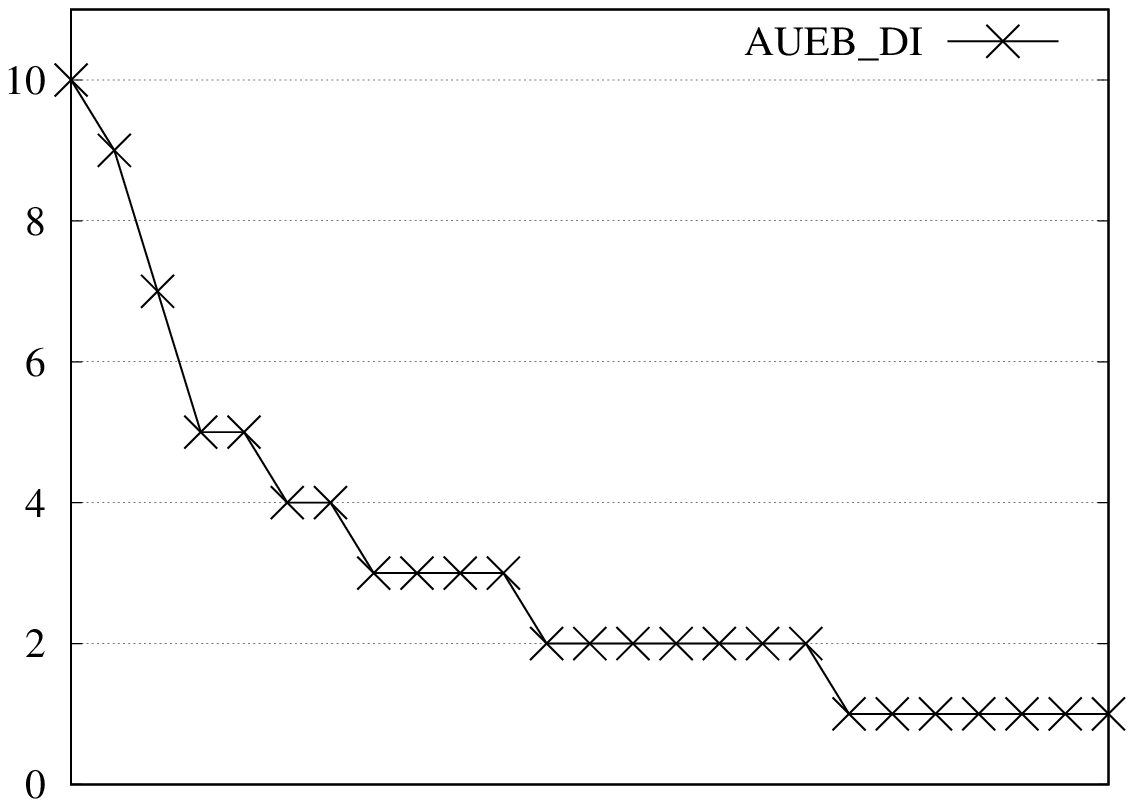}
\includegraphics[scale=.3155]{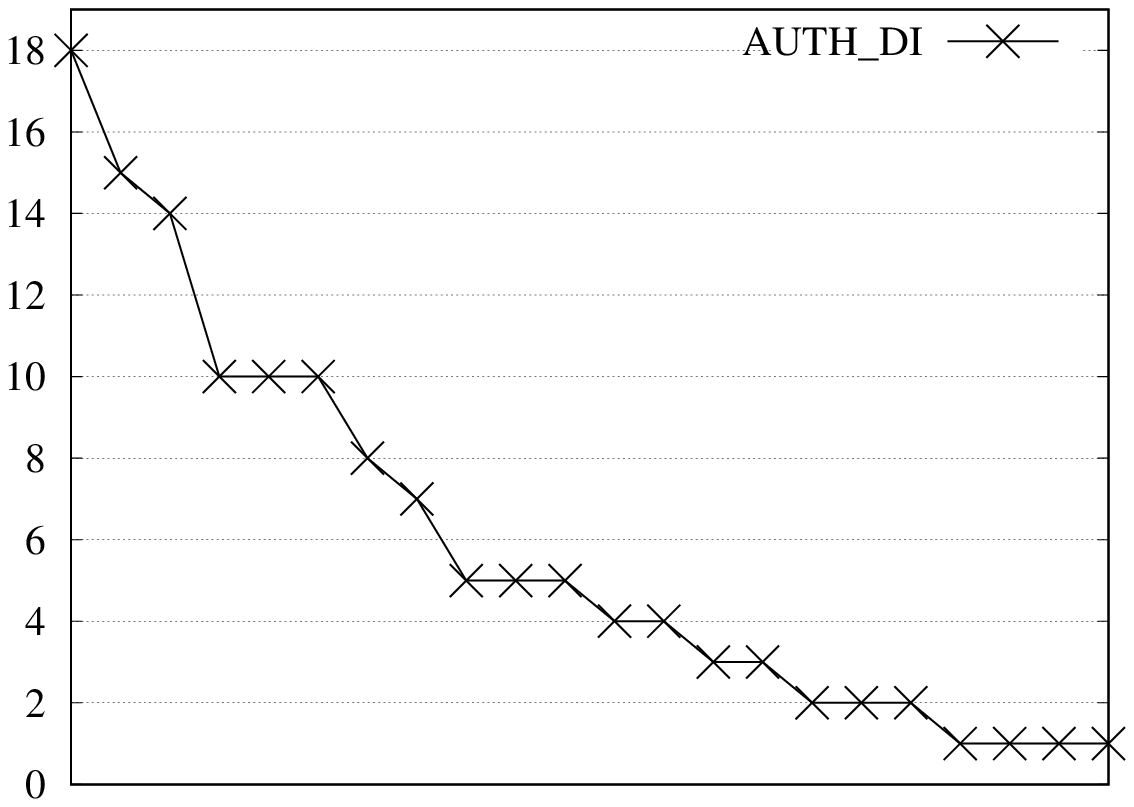}
\includegraphics[scale=.3155]{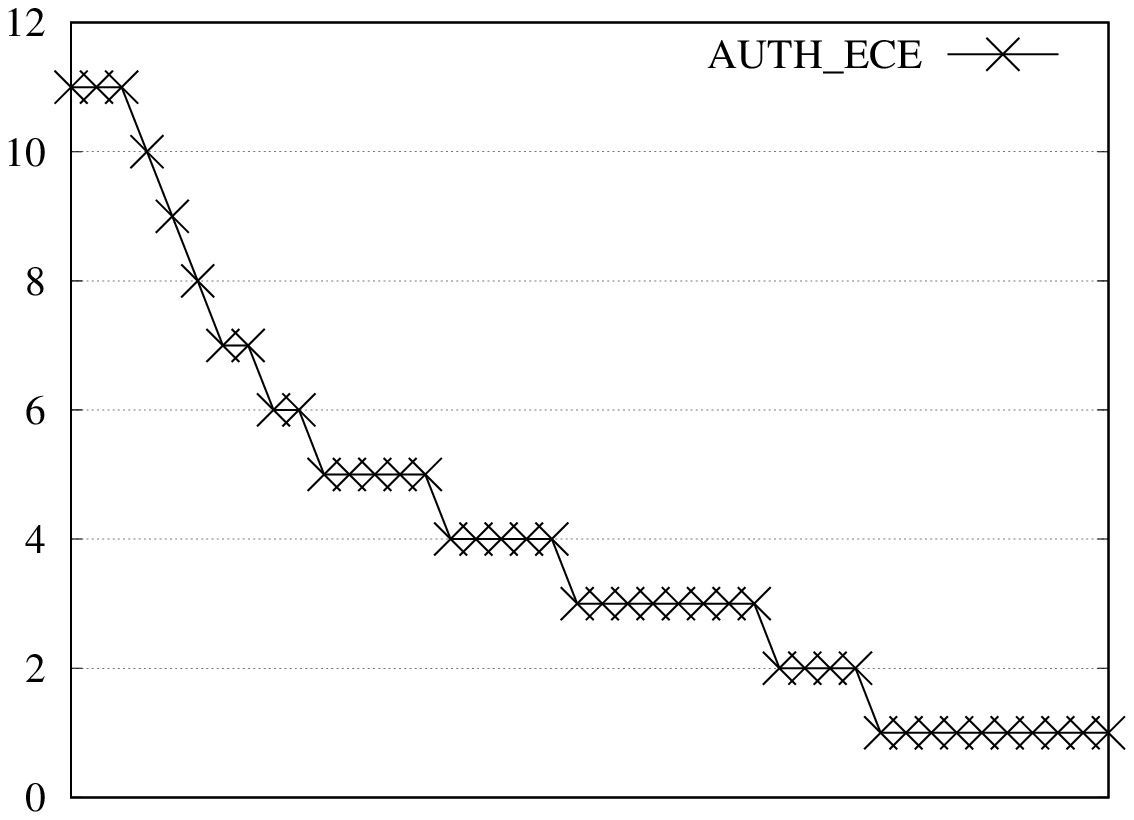}
\includegraphics[scale=.3155]{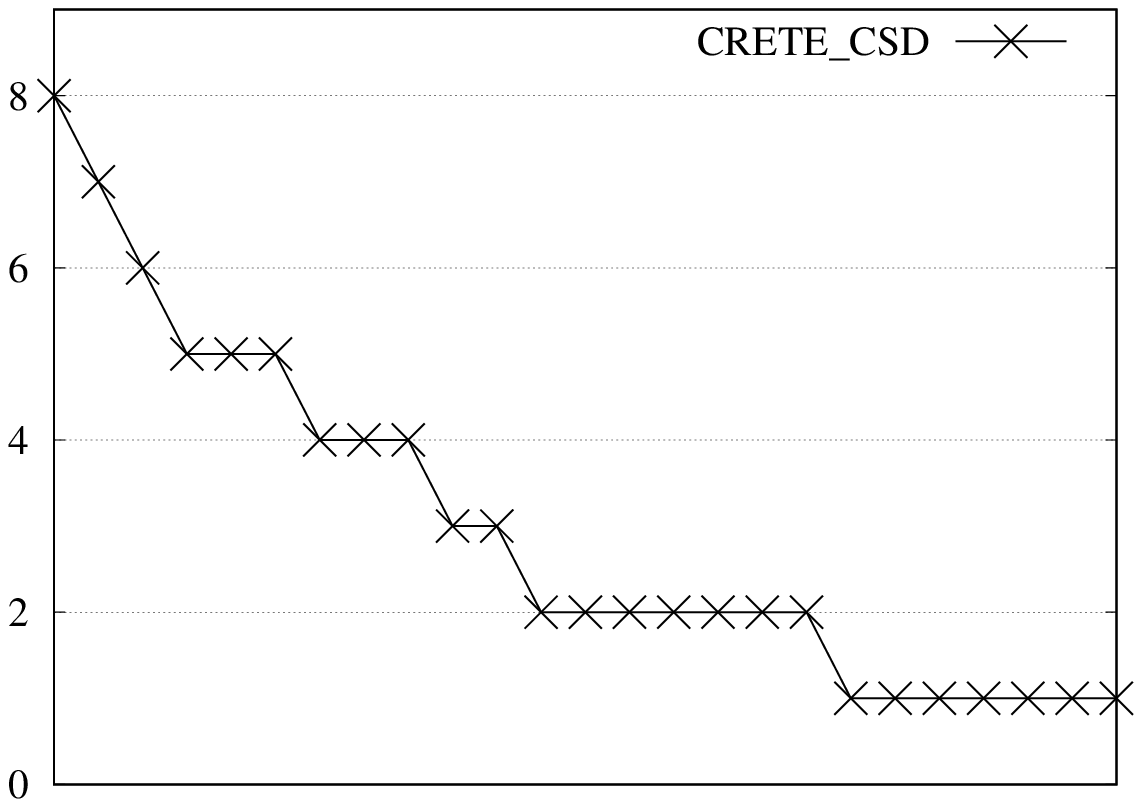}
\includegraphics[scale=.3155]{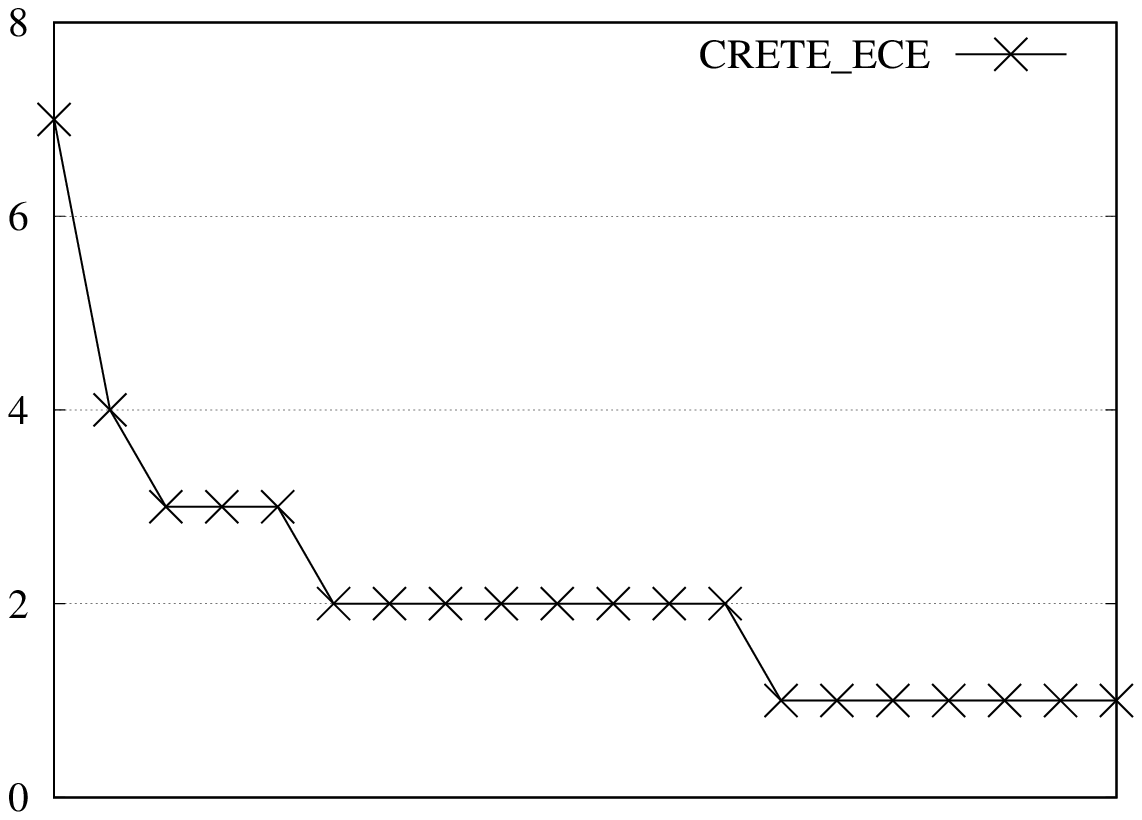}
\includegraphics[scale=.3155]{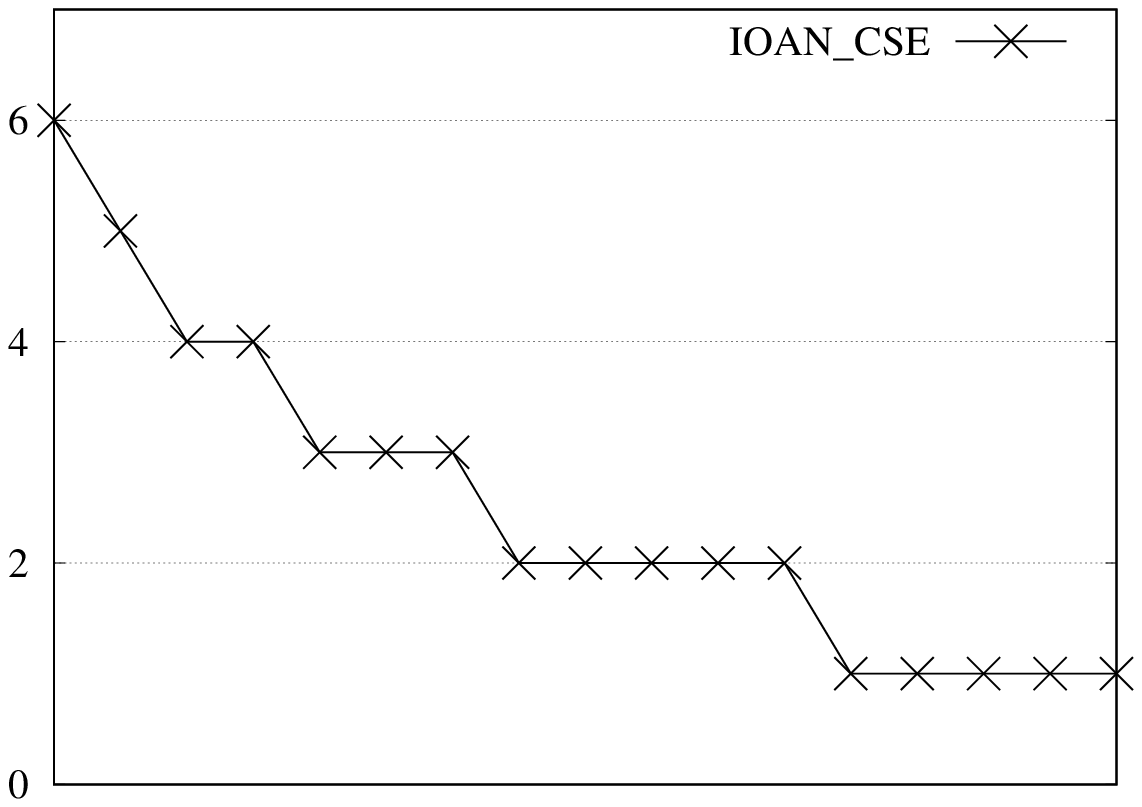}
\includegraphics[scale=.3155]{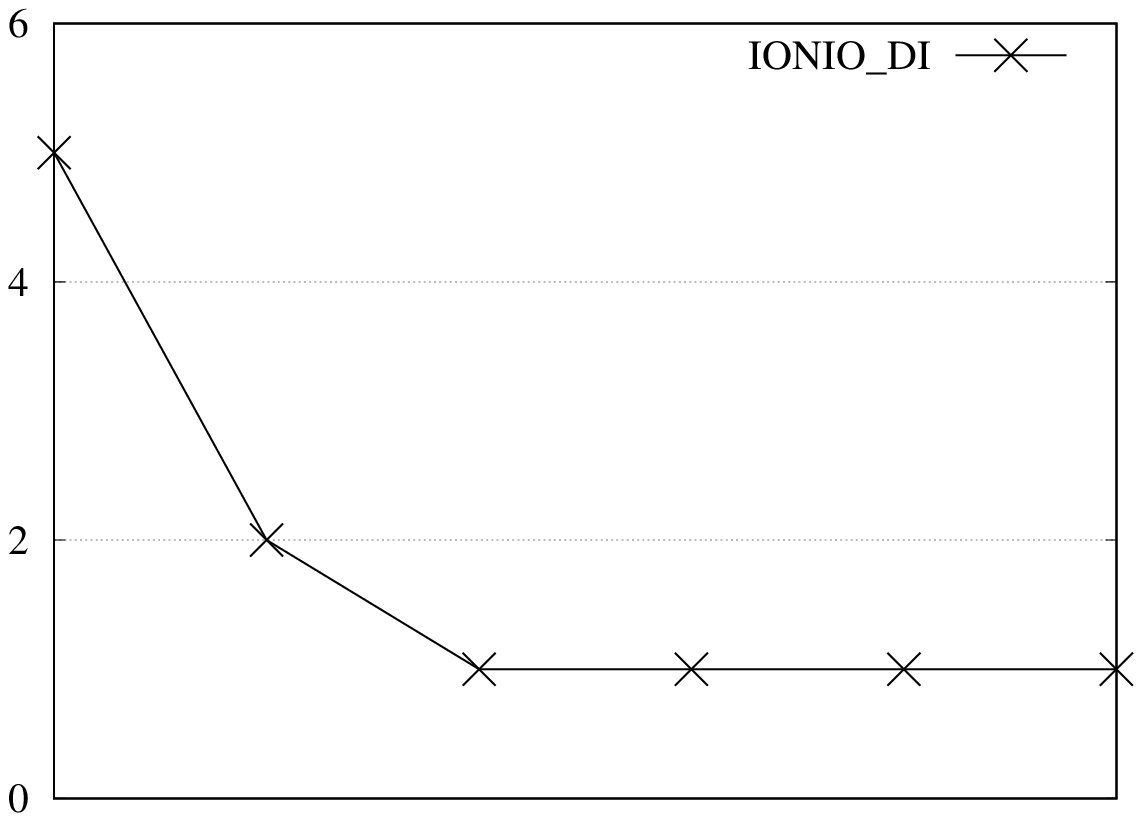}
\includegraphics[scale=.3155]{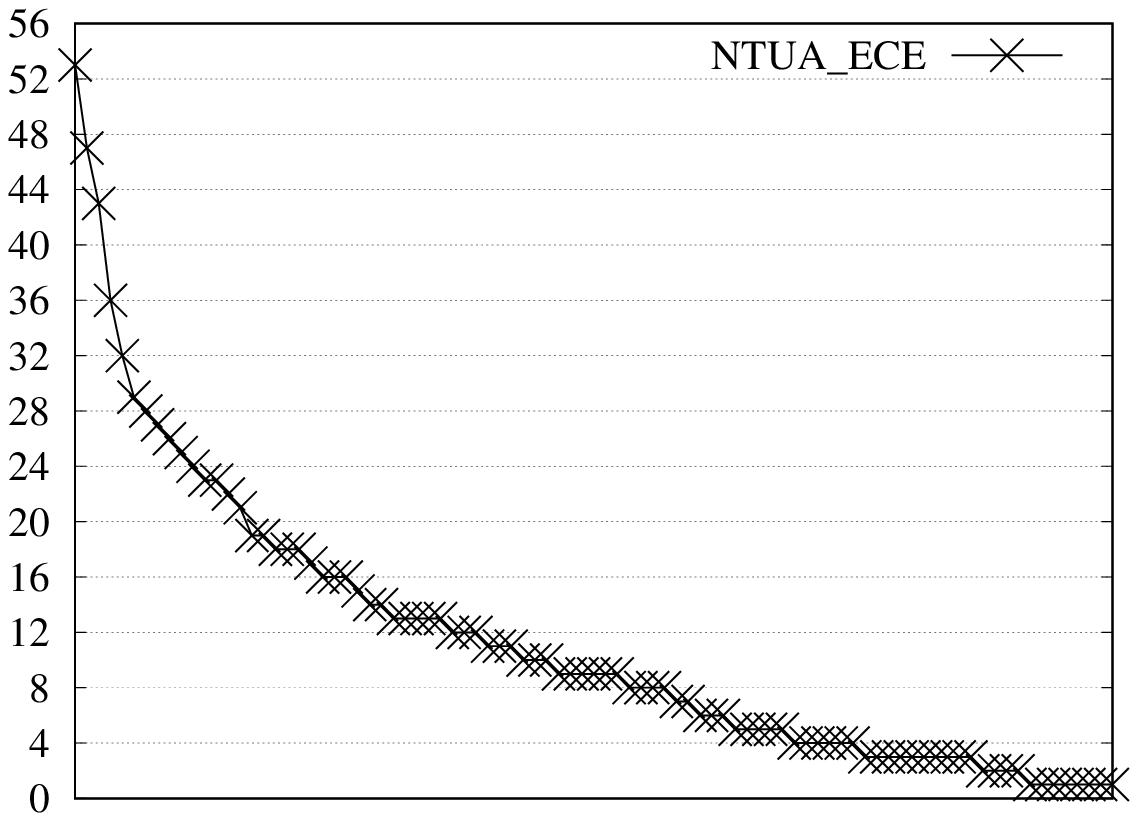}
\includegraphics[scale=.3155]{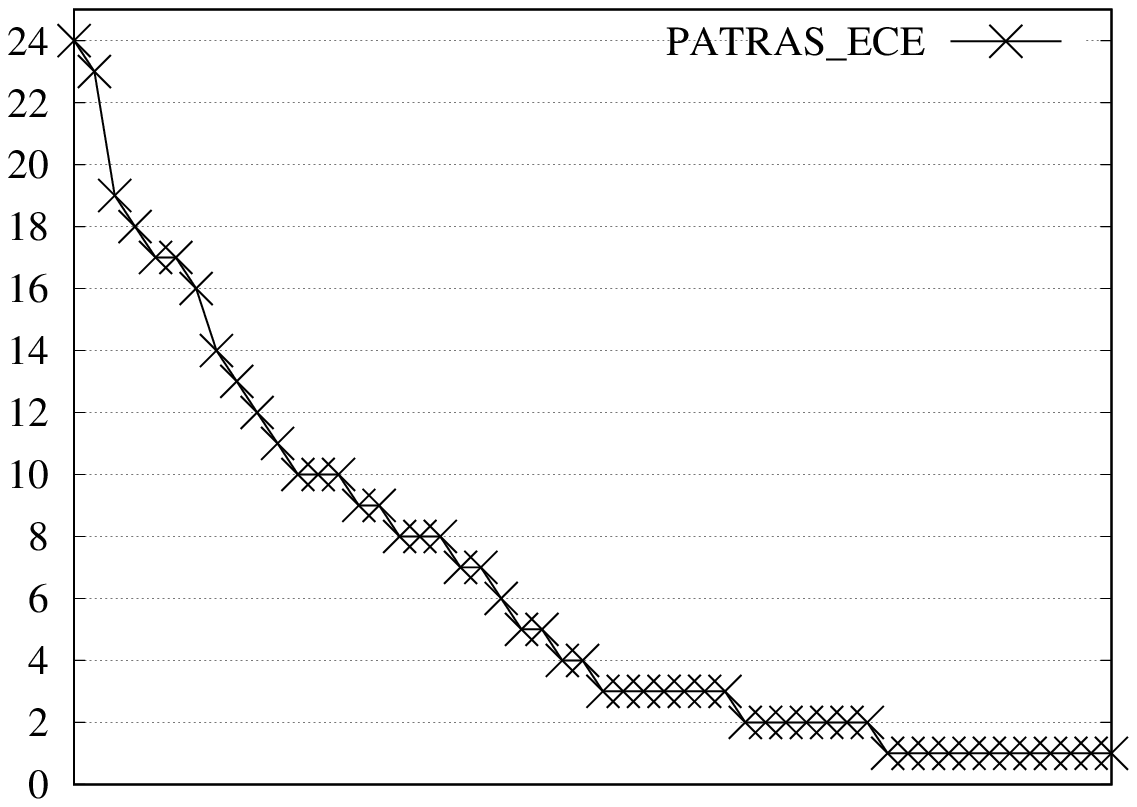}
\includegraphics[scale=.3155]{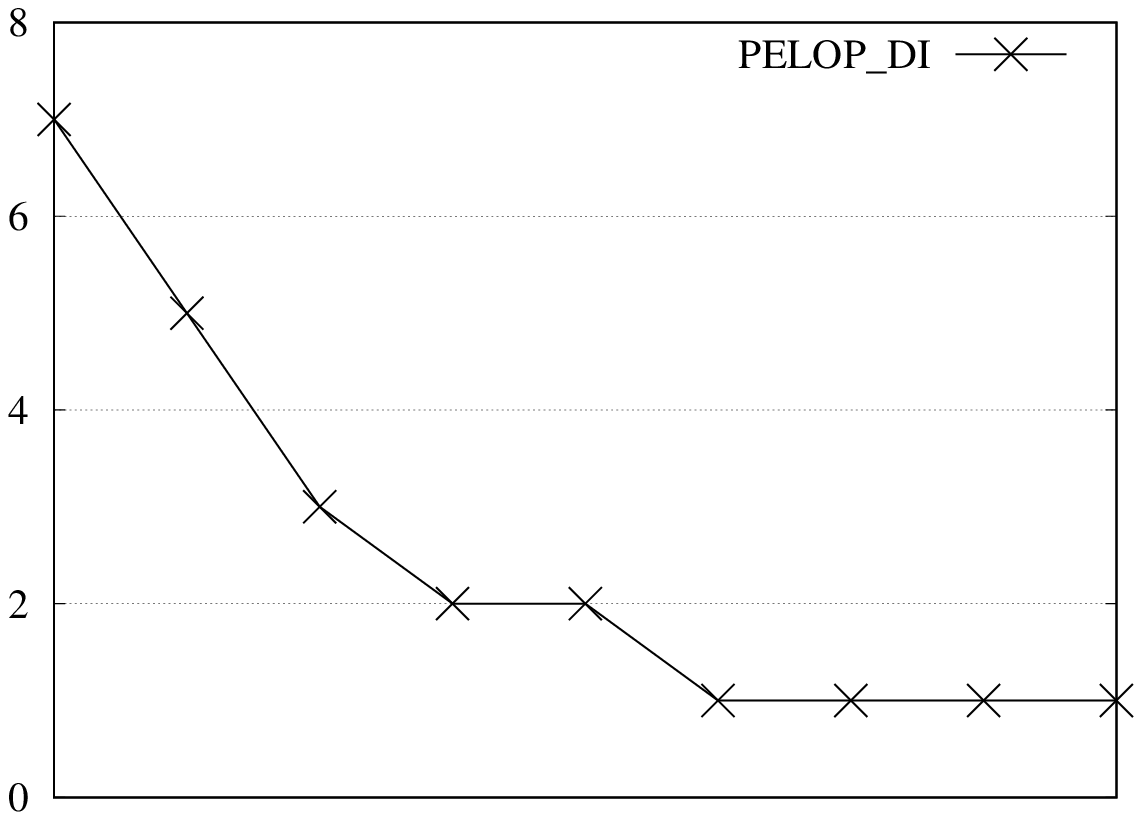}
\includegraphics[scale=.3155]{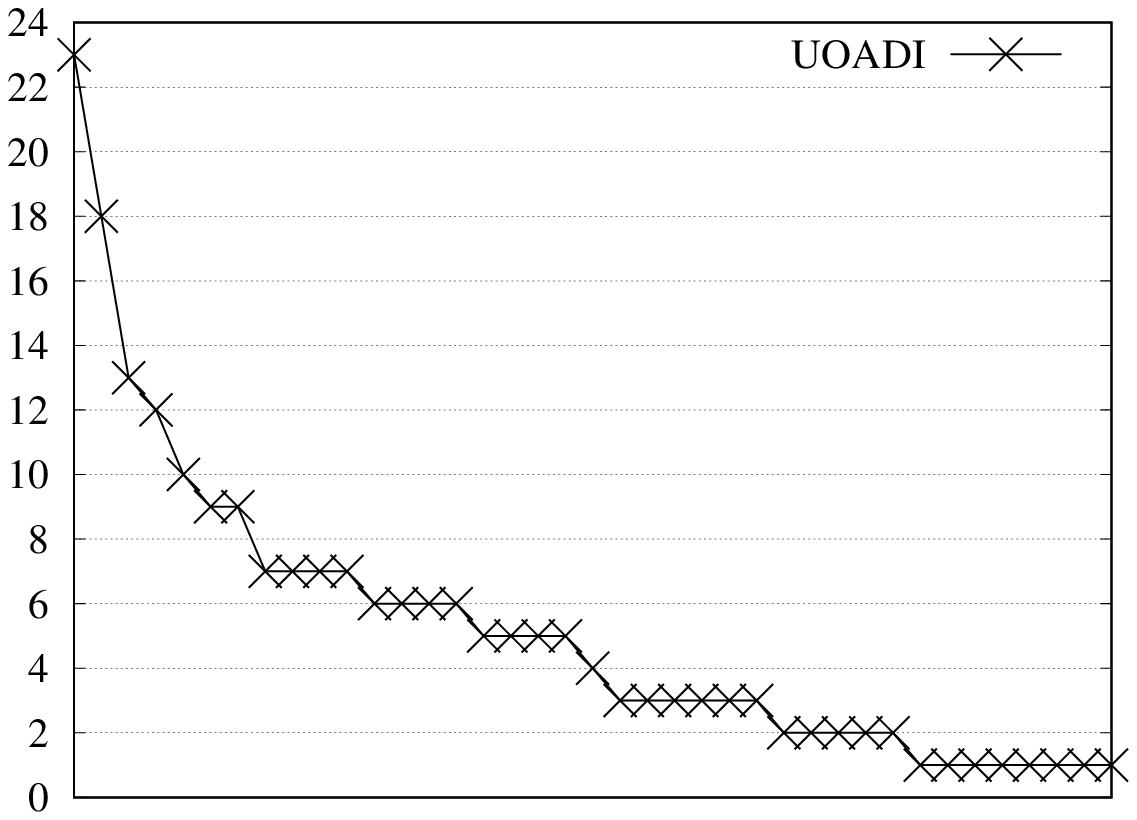}
\includegraphics[scale=.3155]{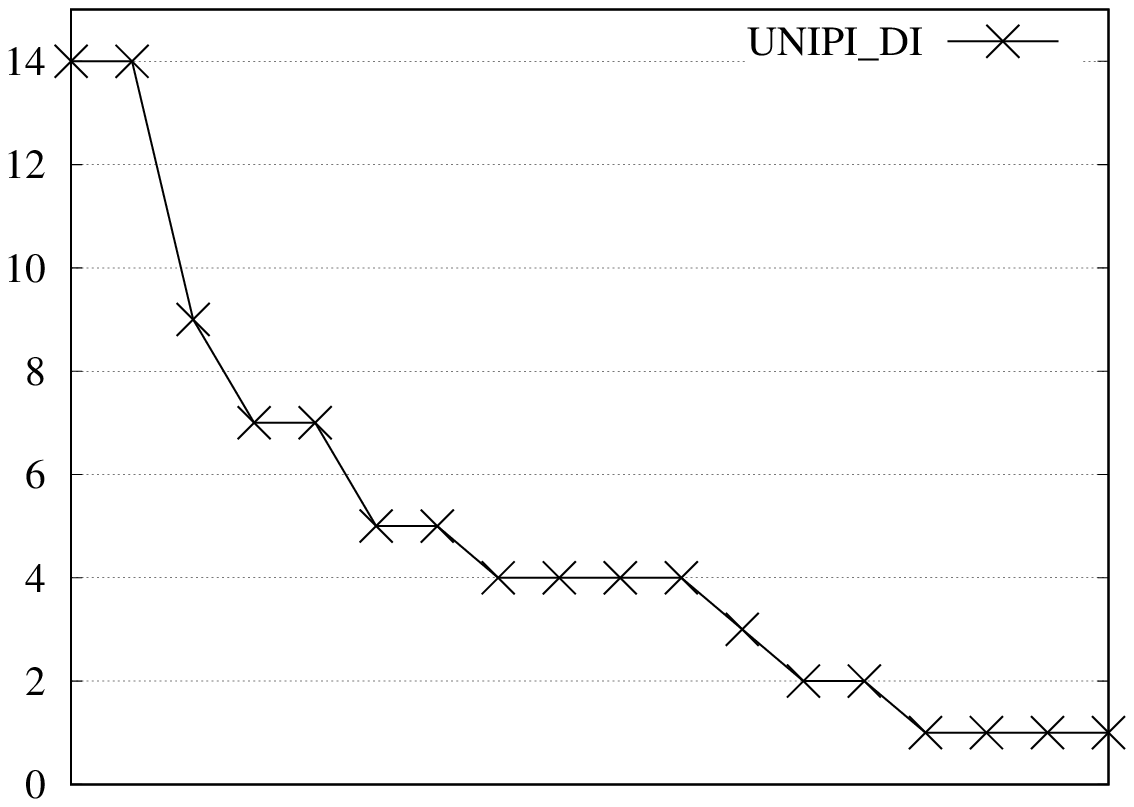}
\includegraphics[scale=.3155]{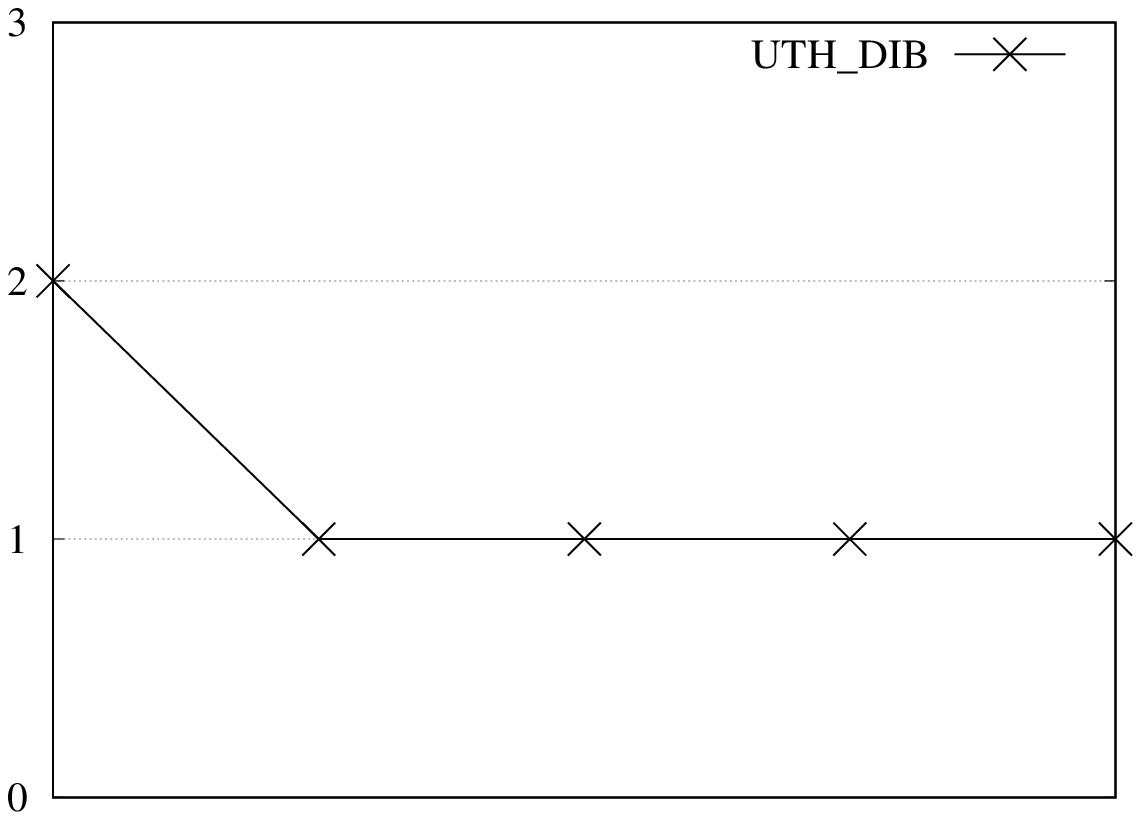}
\includegraphics[scale=.3155]{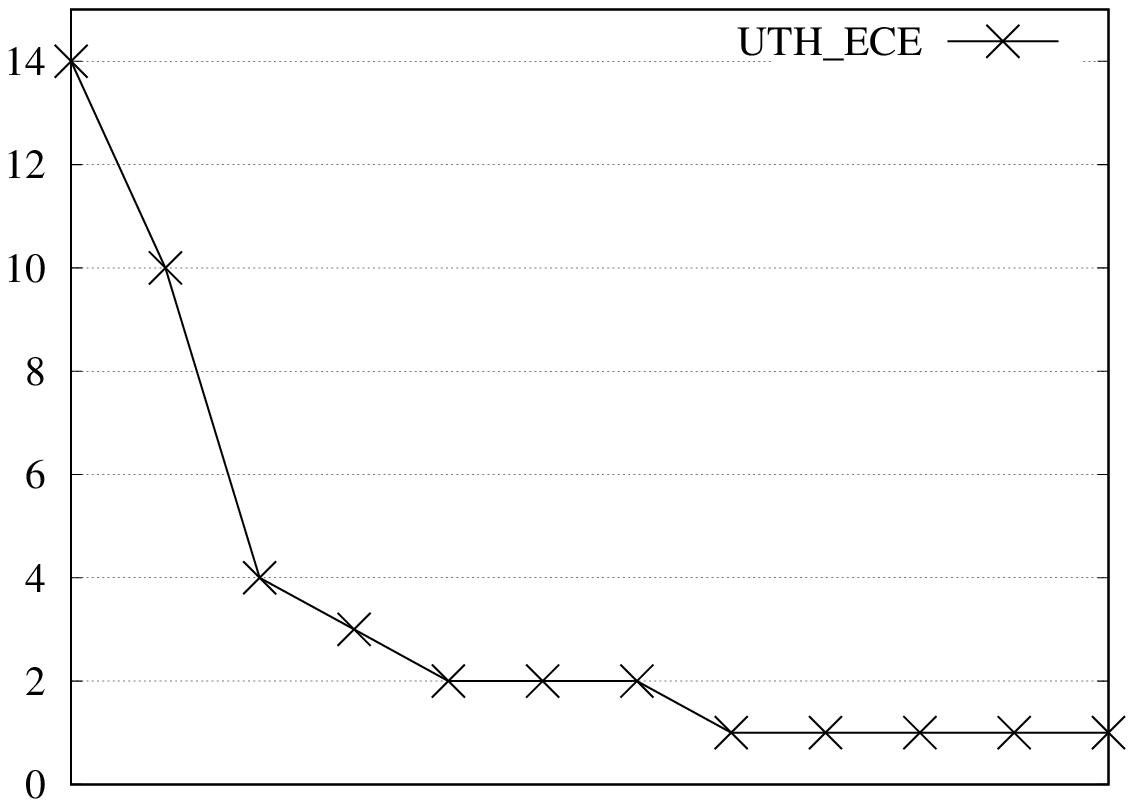}
\end{center}
\vspace*{-1.5\baselineskip}
\caption{Distribution of graduated PhD graduates per supervisor and per department.}
\label{fig-phds-distribution-per-dept}
\end{figure}

In the series of plots in Figure~\ref{fig-phds-distribution-per-dept}, we illustrate the distribution of awarded PhDs per supervisor, after ranking the 
supervisors from the most to the least productive. There are some departments which illustrate a phenomenon where very few supervisors account for the majority 
of awarded PhDs. Such departments are the following: a) AUEB\_DI, where the top-$3$ accounts for more than~$33$\% of all PhDs ($10+9+7=26$ out of~$77$), 
b) AUTH\_DI, where the top-$3$ accounts for more than~$35$\% of all PhDs ($18+15+14=47$ out of~$131$), and c) UTH\_ECE, where the top-$3$ accounts for more 
than~$66$\% of all PhDs ($14+10+4=28$ out of~$43$). Various factors can explain this observation; for instance, the high discrepancy in `scientific quality' 
among the faculty members, and/or the significant discrepancies in `seniority' among faculty members, and/or unequal funding among faculty members. On the other
hand, there exist departments such as CRETE\_CSD, AUTH\_ECE, PATRAS\_ECE, which present a more uniform distribution pattern.

\begin{figure}[!hbt]
\begin{center}
\includegraphics[scale=.6]{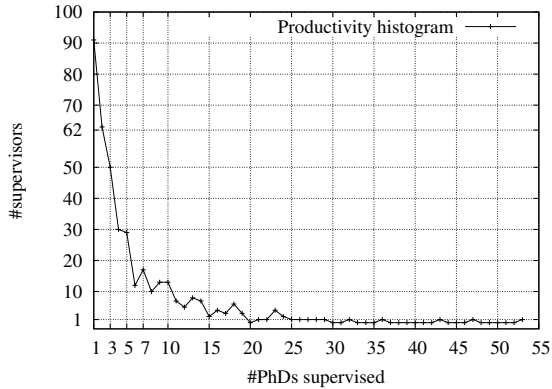}
\end{center}
\vspace*{-\baselineskip}
\caption{Supervisors' productivity: number of supervisors ($y$-axis) as a function of the number of PhD graduates ($x$-axis).}
\label{fig-freq-hist-phds-supervisors}
\end{figure}

Figure~\ref{fig-freq-hist-phds-supervisors} depicts the histogram of supervisors' productivity, i.e., how many supervisors ($y$-axis) have supervised a specific
number ($x$-axis) of PhDs. The curve is highly skewed; we can see that $204$ ($=91 + 63 + 50$) out of $391$ supervisors, i.e., more than $52$\%, have supervised 
at most three PhDs. On the other hand, $18$ supervisors ($4.46$\%) have supervised~$527$ PhDs (22.6\%) -- those at the far right end of $x$-axis (with value 
greater than~$20$).

Before proceeding to show the results about the top- and lowest-impact supervisors, we need to stress that these results {\it do not} necessarily reflect 
the scientific quality of the respective scientists. There are several reasons behind that, e.g., a) the supervisor does not conduct his/her research mainly with 
PhDs, but with post-doctoral researchers, with other faculty members, with research institutions' or industrial personnel, b) the underperformance of the PhD, 
c) his/her research via PhDs has not received attention yet, but it will in the future, and so on.

Concerning the top- and lowest-impact supervisors, we present in Table~\ref{supervisors-top-impact-normalized-per-dept} the top-$3$ -- by total number of 
citations normalized to the number of PhDs (rounded to the closest integer value) for each department, and in 
Table~\ref{supervisors-lowest-impact-normalized-per-dept} the lowest-$3$. 
The selection of top- and lowest-performers was done among those who have supervised at least three PhDs; we feel that this is rational decision. Thus,
we would not show as top-performer, for instance, someone who has supervised, say two PhDs and both of them are in the top-$2$ of a PhDs ranked list.
Similarly, we would show as low-performer someone who has supervised, say three PhDs instead of someone with two PhDs irrespectively of their PhDs' performance.
One could challenge our decision by arguing that it is better to have supervised three PhDs with average or low impact compared to having supervised only one 
with high impact, and so on. 

Nevertheless, our decision is based partly on statistical confidence and partly on our experience while processing these data. This latter factor taught us
that the performance of PhDs of the same supervisor is relatively consistent i.e., for supervisors that had awarded three or more PhDs, we almost never 
observed a situation where a single PhD was responsible for the {\it vast} amount of productivity or impact of that specific supervisor; we encountered
many cases where some PhD was by far the best of that supervisor, but not orders of magnitude better that the rest of the PhDs of that supervisor. 
At this point, we need to emphasize that this is the main reason that we have used `averages' to quantify the impact of a supervisor.\footnote{On the contrary, 
the ranking of universities will be mainly based on median values, instead of means (averages) because of the evident power-law behavior of individual PhD 
performance with respect to the entire set of PhDs of a department (cf.\ Section~\ref{sec-depts-rankings}).}

\begin{table}[!hbt]
\center
\begin{tabular}{||@{}l@{}|@{}c@{}|@{}c@{}|@{}c@{}||}\hline\hline
{\bf Dept}           & {\bf top-1}                          & {\bf top-2}                         & {\bf top-3}\\\hline
{\tiny AEGEAN\_ICSD} & {\scriptsize Kormentzas ($296$)}     & {\scriptsize Gritzalis ($117$)}     & {\scriptsize Kambourakis ($102$)}\\\hline
{\tiny AUEB\_DI}     & {\scriptsize Kiountouzis ($411$)}    & {\scriptsize Vazirgiannis ($369$)}  & {\scriptsize Giannakoudakis ($306$)}\\\hline
{\tiny AUTH\_DI}     & {\scriptsize Pitas ($686$)}          & {\scriptsize Manolopoulos ($518$)}  & {\scriptsize Vlahavas ($477$)}\\\hline
{\tiny AUTH\_ECE}    & {\scriptsize Strintzis ($454$)}      & {\scriptsize Karagiannidis ($332$)} & {\scriptsize Tsiboukis ($298$)}\\\hline
{\tiny CRETE\_CSD}   & {\scriptsize Orphanoudakis ($1074$)} & {\scriptsize Tziritas ($925$)}      & {\scriptsize Konstantopoulos ($425$)}\\\hline
{\tiny CRETE\_ECE}   & {\scriptsize Kalaitzakis ($1055$)}   & {\scriptsize Stavrakakis ($803$)}   & {\scriptsize Zervakis ($250$)}\\\hline
{\tiny IOAN\_CSE}    & {\scriptsize Pitoura ($273$)}        & {\scriptsize Lagaris ($227$)}       & {\scriptsize Lykas ($192$)}\\\hline
{\tiny IONIO\_DI}    & {\scriptsize Vlamos ($99$)}          & ---                                 & ---\\\hline
{\tiny NTUA\_ECE}    & {\scriptsize Papadopoulos ($1060$)}  & {\scriptsize Sellis ($972$)}        & {\scriptsize Hizanidis ($640$)}\\\hline
{\tiny PATRAS\_ECE}  & {\scriptsize Groumpos ($1307$)}      & {\scriptsize King ($689$)}          & {\scriptsize Kotsopoulos ($319$)}\\\hline
{\tiny PELOP\_DI}    & {\scriptsize Simos ($516$)}          & {\scriptsize Vassilakis ($39$)}     & {\scriptsize Vlachos ($18$)}\\\hline
{\tiny UA\_DI}       & {\scriptsize Ioannidis ($411$)}      & {\scriptsize Syvridis ($410$)}      & {\scriptsize Paschalis ($293$)}\\\hline
{\tiny UNIPI\_DI}    & {\scriptsize Theodoridis ($294$)}    & {\scriptsize Virvou ($157$)}        & {\scriptsize Alexandris ($107$)}\\\hline
{\tiny UTH\_DIB}     & ---                                  & ---                                 & ---\\\hline
{\tiny UTH\_ECE}     & {\scriptsize Houstis E. ($270$)}     & {\scriptsize Tassiulas ($233$)}     & {\scriptsize Lalis ($108$)}\\\hline\hline
\end{tabular}
\vspace*{.05\baselineskip}
\caption{Top-$3$ impactful supervisors per department. 
The number in parenthesis is the average number of citations received by his/her PhDs.}
\label{supervisors-top-impact-normalized-per-dept}
\end{table}

A consequence of our decision to consider the value of three supervised PhDs as the threshold to include a supervisor into 
Tables~\ref{supervisors-top-impact-normalized-per-dept} and~\ref{supervisors-lowest-impact-normalized-per-dept} is that the former table has no entries for 
UTH\_DIB (no supervisors has supervised more than two PhDs), and partial entries for IONIO\_DI (only one supervisor has supervised more than two). Additionally, 
there are empty entries in Table~\ref{supervisors-lowest-impact-normalized-per-dept} in all cases where a department has less than~$6$ supervisors with more than 
three supervised PhDs, because we have used three of them as entries in Table~\ref{supervisors-top-impact-normalized-per-dept}.

\begin{table}[!bht]
\center
\begin{tabular}{||@{}l|@{}c@{}|@{}c@{}|@{}c@{}||}\hline\hline
{\bf Department}           & {\bf bottom-1}                          & {\bf bottom-2}                      & {\bf bottom-3}\\\hline
{\scriptsize AEGEAN\_ICSD} & {\scriptsize Loukis ($21$)}             & ---                                 & ---\\\hline
{\scriptsize AUEB\_DI}     & {\scriptsize Sideri ($16$)}             & {\scriptsize Mageirou ($59$)}       & {\scriptsize Apostolopoulos ($99$)}\\\hline
{\scriptsize AUTH\_DI}     & {\scriptsize Demetriadis ($81$)}        & {\scriptsize Miliou ($85$)}         & {\scriptsize Lazos ($95$)}\\\hline
{\scriptsize AUTH\_ECE}    & {\scriptsize Dokouzyannis ($4$)}        & {\scriptsize Chrissoulidis ($20$)}  & {\scriptsize Xenos ($36$)}\\\hline
{\scriptsize CRETE\_CSD}   & {\scriptsize Stephanidis ($21$)}        & {\scriptsize Katevenis ($137$)}     & {\scriptsize Stylianou ($138$)}\\\hline
{\scriptsize CRETE\_ECE}   & {\scriptsize Dollas ($79$)}             & {\scriptsize Papaefstathiou ($86$)} & ---\\\hline
{\scriptsize IOAN\_CSE}    & {\scriptsize Kondis ($34$)}             & {\scriptsize Fudos ($35$)}          & ---\\\hline
{\scriptsize IONIO\_DI}    & ---                                     & ---                                 & ---\\\hline
{\scriptsize NTUA\_ECE}    & {\scriptsize Ioannides Maria ($14$)}    & {\scriptsize Bourkas ($17$)}        & {\scriptsize Matsopoulos ($26$)}\\\hline
{\scriptsize PATRAS\_ECE}  & {\scriptsize Denazis ($31$)}            & {\scriptsize Tsanakas D.~K.($38$)}  & {\scriptsize Paliouras ($41$)}\\\hline
{\scriptsize PELOP\_DI}    & ---                                     & ---                                 & ---\\\hline
{\scriptsize UA\_DI}       & {\scriptsize Rondogiannis ($27$)}       & {\scriptsize Varoutas ($28$)}       & {\scriptsize Manolakos ($31$)}\\\hline
{\scriptsize UINPI\_PI}    & {\scriptsize Panagiotopoulos, T. ($3$)} & {\scriptsize Siskos ($4$)}          & {\scriptsize Foundas ($6$)}\\\hline
{\scriptsize UTH\_DIB}     & ---                                     & ---                                 & ---\\\hline
{\scriptsize UTH\_ECE}     & {\scriptsize Stamoulis G.~I.($15$)}     & ---                                 & ---\\\hline\hline
\end{tabular}
\vspace*{.05\baselineskip}
\caption{Bottom-$3$ impactful supervisors per department, normalized per number of supervised PhDs. 
The number in parenthesis is the average number of citations received by his/her PhDs.}
\label{supervisors-lowest-impact-normalized-per-dept}
\end{table}

Contrasting Table~\ref{supervisors-top-impact-normalized-per-dept} to Table~\ref{tab-supervisors-productivity-per-dept}, we observe significant differences for 
many departments with respect to: a) the identity of supervisors comprising the tables, and b) the relative ranking of supervisors. This is not a surprising 
result however. The reader can make his/her own detailed observations.

As mentioned before, Table~\ref{supervisors-lowest-impact-normalized-per-dept} illustrates the lowest-$3$ supervisors impact-wise per department. We can see that 
even supervisors who have appeared in Table~\ref{tab-supervisors-productivity-per-dept}, are included in this table as well. A very interesting result (which 
will be explained in Section~\ref{sec-depts-rankings}) concerns the significantly higher performance of the low-performing supervisors coming from CRETE\_CSD, 
CRETE\_ECE and AUTH\_DI; these supervisors have four to five times the performance of their colleagues appearing in 
Table~\ref{supervisors-lowest-impact-normalized-per-dept}.

Finally, for completeness purposes we include Table~\ref{tab-bottom-10-impactfull-supervisors} to show the worse performing supervisors impact-wise. It seems
that UNIPI\_DI appears quite frequently on this list.

\begin{table}[!hbt]
\center
\begin{tabular}{||c|c|c||}\hline\hline
{\bf avg {\#}citations per PhD} & {\bf supervisor}                  & {\bf Department}\\\hline
$3$                             & Panagiotopoulos, T.               & UNIPI\_DI\\\hline
$4$                             & \minitab[c]{Dokouzyannis\\Siskos} & \minitab[c]{AUTH\_ECE\\UNIPI\_DI}\\\hline
$6$                             & Foundas                           & UNIPI\_DI\\\hline
$11$                            & Assimakopoulos                    & UNIPI\_DI\\\hline
$14$                            & Ioannides Maria                   & NTUA\_ECE\\\hline
$15$                            & Stamoulis G.~I.                   & UTH\_ECE\\\hline
$16$                            & Sideri                            & AUEB\_DI\\\hline
$17$                            & Bourkas                           & NTUA\_ECE\\\hline
$18$                            & Vlachos                           & PEL\_DI\\\hline
$20$                            & Chrissoulidis                     & AUTH\_ECE\\\hline
\end{tabular}
\vspace*{.05\baselineskip}
\caption{Bottom-$10$ impactful supervisors across all departments.}
\label{tab-bottom-10-impactfull-supervisors}
\end{table}

\section{Performance of PhDs}
\label{sec-phds-performance}

As explained in Section~\ref{sec-dept-aggr-stats}, impact and graduation year are not strongly correlated, thus we refrain to normalize the results based on 
time-based parameters. Normalizing by or showing the graduation year is not significant for one more reason; the duration of PhD studies of each PhD is not 
available to us, and simply normalizing by the year of the first published article will in principle bias the results towards those who were not productive 
(impactful) during the first year(s) and did all their work during the final year(s) of their PhD studies.

In the $h$-index based ranking, we show only those PhDs with $h$-index value greater than or equal to~$10$ (or the highest ranked PhD only, if there is none 
with $h$-index $\geq 10$), because of the too many ties at lower values, and thus the low discriminative power of $h$-index at these levels of performance.

Tables~\ref{tab-top-10-PhDs-AEGEAN-ICSD}--\ref{tab-top-10-PhDs-UTH-ECE} present for each department the top-$10$ ranking of the PhDs based on productivity and 
impact. We are not going to describe in details the table entries for each department, but will only summarize some generic observations.
\begin{itemize}
\item Many of the PhDs which made it into top-$10$ are now faculty members either in their department or other departments in national or international 
      universities. One could argue that their current position explains their successful statistics, or that their statistics were important for their 
      recruitment; we tend to believe that they are both valid explanations. 
\item Productivity and impact are not correlated, and thus many PhDs ranked high in terms of total number of published articles are found in lower positions when
      looking at the impact-based ranking.
\item Apart from a couple of PhDs with outstanding performance (e.g., Dimitris Papadias, NTUA\_ECE, now professor at HKUST), we do not recognize substantial 
      differences among the top-performing PhDs relative to impact. This is a qualitative observation, however, that needs to be verified quantitatively in a 
      future work.
\item Some departments exhibit far stronger {\it inbreeding} than others.\footnote{We do not present the list of faculty members here, but this information is 
      readily available on the Internet.} The analysis of the career paths of PhDs is a very interesting topic on itself. It would be interesting to record 
      percentages of PhDs employed by national or international industries, by national governmental organizations; what percentage of PhDs becomes faculty member
      in its own department, or in other national or international departments, but this is a separate scientometric task beyond the scope of the current article.
\end{itemize}

\begin{table}[!hbt]
\center
\begin{tabular}{||c|c|c||}\hline\hline
{\bf {\#}articles}                                                                                   & {\bf {\#}citations}               & $h$-{\bf index}       \\\hline
{\scriptsize Mastorakis ($105$)}                                                                     & {\scriptsize Kambourakis ($919$)} & {\scriptsize Kambourakis ($18$)} \\\hline
{\scriptsize Kambourakis ($103$)}                                                                    & {\scriptsize Mastorakis ($504$)}  & {\scriptsize Mastorakis ($12$)} \\\hline
{\scriptsize Belsis ($50$)}                                                                          & {\scriptsize Kostis ($424$)}      & {\scriptsize Bourdena ($11$)} \\\hline
{\scriptsize Bourdena ($42$)}                                                                        & {\scriptsize Bourdena ($307$)}    & {\scriptsize Kostis ($10$)} \\\hline
{\scriptsize Vassis ($28$)}                                                                          & {\scriptsize Pliakas ($197$)}     & \multirow{6}*{}\\\cline{1-2}
{\scriptsize Kostis ($24$)}                                                                          & {\scriptsize Vassis ($179$)}      & \\\cline{1-2}
\minitab[c]{{\scriptsize Makris ($23$)}\\{\scriptsize Miritzis ($23$)}\\{\scriptsize Tsohou ($23$)}} & {\scriptsize Damopoulos ($144$)}  & \\\cline{1-2}
{\scriptsize Kollias ($21$)}                                                                         & {\scriptsize Makris ($136$)}      & \\\cline{1-2}
{\scriptsize Karopoulos ($19$)}                                                                      & {\scriptsize Kolias ($128$)}      & \\\cline{1-2}
{\scriptsize Pliakas ($18$)}                                                                         & {\scriptsize Tsohou ($120$)}      & \\\cline{1-2}\hline\hline
\end{tabular}
\vspace*{.05\baselineskip}
\caption{Top-$10$ PhDs of the AEGEAN\_ICSD department.}
\label{tab-top-10-PhDs-AEGEAN-ICSD}
\end{table}

\begin{table}[!hbt]
\center
\begin{tabular}{||@{}c@{}|@{}c@{}|@{}c@{}||}\hline\hline
{\bf {\#}articles}                                                            & {\bf {\#}citations}               & $h$-{\bf index} \\\hline
{\scriptsize Tryfonas ($93$)}                                                 & {\scriptsize Halkidi  ($2134$)}   & {\scriptsize Doulkeridis ($17$)} \\\hline
{\scriptsize Varlamis ($67$)}                                                 & {\scriptsize Eirinaki ($911$)}    & {\scriptsize Vlachou ($16$)} \\\hline
{\scriptsize Doulkeridis ($64$)}                                              & {\scriptsize Doulkeridis ($901$)} & {\scriptsize Papadakis ($15$)} \\\hline
{\scriptsize Oikonomou ($55$)}                                                & {\scriptsize Vlachou ($764$)}     & \minitab[c]{{\scriptsize Eirinaki ($13$)}\\{\scriptsize Papaioannou ($13$)}\\{\scriptsize Tryfonas ($13$)}} \\\hline
\minitab[c]{{\scriptsize Papadakis ($53$)}\\{\scriptsize Papaioannou ($53$)}} & {\scriptsize Papadakis ($541$)}   & \minitab[c]{{\scriptsize Halkidi ($11$)}\\{\scriptsize Kokolakis ($11$)}\\{\scriptsize Oikonomou ($11$)}} \\\hline
{\scriptsize Vlachou ($50$)}                                                  & {\scriptsize Varlamis ($515$)}    & \minitab[c]{{\scriptsize Antoniadis ($10$)}\\{\scriptsize Karyda ($10$)}\\{\scriptsize Theocharidou ($10$)}\\{\scriptsize Tsatsaronis ($10$)}\\{\scriptsize Varlamis ($10$)}} \\\hline
{\scriptsize Kokolakis ($46$)}                                                & {\scriptsize Papaioannou ($489$)} & \multirow{4}*{}\\\cline{1-2}
\minitab[c]{{\scriptsize Eirinaki ($37$)}\\{\scriptsize Theocharidou ($37$)}} & {\scriptsize Ververidis ($470$)}  & \\\cline{1-2}
{\scriptsize Fotiou ($35$)}                                                   & {\scriptsize Kokolakis ($466$)}   & \\\cline{1-2}
{\scriptsize Frangoudis ($33$)}                                               & {\scriptsize Fotiou ($437$)}      & \\\cline{1-2}\hline\hline
\end{tabular}
\vspace*{.05\baselineskip}
\caption{Top-$10$ PhDs of the AUEB\_DI department.}
\label{tab-top-10-PhDs-AUEB-DI}
\end{table}

\begin{table}[!hbt]
\center
\begin{tabular}{||@{}c@{}|@{}c@{}|@{}c@{}||}\hline\hline
{\bf {\#}articles}                                                                     & {\bf {\#}citations}                   & $h$-{\bf index} \\\hline
{\scriptsize Tefas ($245$)}                                                            & {\scriptsize Tsoumakas ($2744$)}      & {\scriptsize Tefas ($26$)} \\\hline
{\scriptsize Vakali ($169$)}                                                           & {\scriptsize Zafeiriou ($2405$)}      & {\scriptsize Zafeiriou ($25$)} \\\hline
{\scriptsize Zafeiriou ($152$)}                                                        & {\scriptsize Tefas ($2363$)}          & {\scriptsize Nanopoulos ($23$)} \\\hline
{\scriptsize Bassiliades ($150$)}                                                      & {\scriptsize Nanopoulos ($2093$)}     & {\scriptsize Tsoumakas ($21$)} \\\hline
{\scriptsize Nanopoulos ($138$)}                                                       & {\scriptsize Vakali ($1981$)}         & {\scriptsize Katsaros, D. ($19$)} \\\hline
{\scriptsize Mavromoustakis ($129$)}                                                   & {\scriptsize Katsaros D. ($1767$)}    & \minitab[c]{{\scriptsize Bassiliades ($18$)}\\{\scriptsize Bors ($18$)}\\{\scriptsize Chatzigeorgiou ($18$)}\\{\scriptsize Vakali ($18$)}} \\\hline
{\scriptsize Nicopolitidis ($126$)}                                                    & {\scriptsize Pallis ($1338$)}         & \minitab[c]{{\scriptsize Pallis ($15$)}\\{\scriptsize Papadopoulos, A. ($15$)}\\{\scriptsize Symeonidis ($15$)}} \\\hline
\minitab[c]{{\scriptsize Demetriadis ($104$)}\\{\scriptsize Papadopoulos, S. ($104$)}} & {\scriptsize Chatzigeorgiou ($1200$)} & {\scriptsize Iosifidis ($14$)} \\\hline
\minitab[c]{{\scriptsize Bors ($101$)}\\{\scriptsize Chatzigeorgiou ($101$)}}          & {\scriptsize Katakis ($1189$)}        & \minitab[c]{{\scriptsize Mavromoustakis ($13$)}\\{\scriptsize Nicopolitidis ($13$)}\\{\scriptsize Papadopoulos, S. ($13$)}\\{\scriptsize Solachidis ($13$)}} \\\hline
{\scriptsize Katsaros, D. ($93$)}                                                      & {\scriptsize Bassiliades ($1164$)}    & \minitab[c]{{\scriptsize Demetriadis ($12$)}\\{\scriptsize Tsekeridou ($12$)}\\{\scriptsize Ververidis ($12$)}} \\\hline
\end{tabular}
\vspace*{.05\baselineskip}
\caption{Top-$10$ PhDs of the AUTH\_DI department.}
\label{tab-top-10-PhDs-AUTH-DI}
\end{table}

\begin{table}[!hbt]
\center
\begin{tabular}{||@{}c@{}|c|c@{}||}\hline\hline
{\bf {\#}articles}                    & {\bf {\#}citations}                  & $h$-{\bf index}     \\\hline
{\scriptsize Mezaris ($167$)}         & {\scriptsize Michalopoulos ($1453$)} & {\scriptsize Michalopoulos ($22$)} \\\hline
{\scriptsize Daras ($151$)}           & {\scriptsize Mezaris ($1366$)}       & {\scriptsize Sounas ($20$)} \\\hline
{\scriptsize Sounas ($108$)}          & {\scriptsize Sounas ($1220$)}        & {\scriptsize Mezaris ($17$)} \\\hline
{\scriptsize Moustakas ($103$)}       & {\scriptsize Daras ($1050$)}         & \minitab[c]{{\scriptsize Daras ($15$)}\\{\scriptsize Zografopoulos ($15$)}} \\\hline
{\scriptsize Athanasiadis ($84$)}     & {\scriptsize Athanasiadis ($817$)}   & \minitab[c]{{\scriptsize Athanasiadis ($14$)}\\{\scriptsize Moustakas ($14$)}} \\\hline
{\scriptsize Papadopoulos, T. ($72$)} & {\scriptsize Bechlioudis ($789$)}    & \minitab[c]{{\scriptsize Chatzidiamantis ($12$)}\\{\scriptsize Thomos ($12$)}} \\\hline
{\scriptsize Zografopoulos ($71$)}    & {\scriptsize Moustakas ($748$)}      & {\scriptsize Polimeridis ($11$)} \\\hline
{\scriptsize Michalopoulos ($69$)}    & {\scriptsize Zografopoulos ($611$)}  & \minitab[c]{{\scriptsize Kosmidou ($10$)}\\{\scriptsize Lioumpas ($10$)}\\{\scriptsize Papadopoulos, T ($10$)}\\{\scriptsize Tsilipakos ($10$)}} \\\hline
{\scriptsize Polimeridis ($55$)}      & {\scriptsize Thomos ($560$)}         & \multirow{2}*{} \\\cline{1-2}
{\scriptsize Zygiridis ($53$)}        & {\scriptsize Zoumas ($549$)}         & \\\cline{1-2}\hline\hline
\end{tabular}
\vspace*{.05\baselineskip}
\caption{Top-$10$ PhDs of the AUTH\_ECE department.}
\label{tab-top-10-PhDs-AUTH-ECE}
\end{table}

\begin{table}[!hbt]
\center
\begin{tabular}{||c|c|c||}\hline\hline
{\bf {\#}articles}                 & {\bf {\#}citations}                 & $h$-{\bf index}     \\\hline
{\scriptsize Siris ($118$)}        & {\scriptsize Komodakis ($2194$)}    & {\scriptsize Komodakis ($22$)} \\\hline
{\scriptsize Spanoudakis ($105$)}  & {\scriptsize Argyros ($2100$)}      & {\scriptsize Argyros ($19$)} \\\hline
{\scriptsize Argyros ($101$)}      & {\scriptsize Petrakis ($1617$)}     & {\scriptsize Spanoudakis ($18$)} \\\hline
{\scriptsize Tzitzikas ($98$)}     & {\scriptsize Siris ($1427$)}        & \minitab[c]{{\scriptsize Lourakis ($17$)}\\{\scriptsize Siris ($17$)}} \\\hline
{\scriptsize Petrakis ($93$)}      & {\scriptsize Spanoudakis ($1210$)}  & {\scriptsize Polychronakis ($16$)} \\\hline
{\scriptsize Zaboulis ($86$)}      & {\scriptsize Lourakis ($1068$)}     & {\scriptsize Petrakis ($15$)} \\\hline
{\scriptsize Komodakis ($83$)}     & {\scriptsize Polychronakis ($879$)} & {\scriptsize Antonatos ($14$)} \\\hline
{\scriptsize Pappas ($73$)}        & {\scriptsize Oikonomidis ($847$)}   & {\scriptsize Zaboulis ($13$)} \\\hline
{\scriptsize Bikakis ($68$)}       & {\scriptsize Antonatos ($683$)}     & {\scriptsize Flouris ($12$)} \\\hline
{\scriptsize Polychronakis ($66$)} & {\scriptsize Kyriazis ($665$)}      & \minitab[c]{{\scriptsize Kritikos ($11$)}\\{\scriptsize Panagiotakis ($11$)}\\{\scriptsize Tzitzikas ($11$)}} \\\hline
\end{tabular}
\vspace*{.05\baselineskip}
\caption{Top-$10$ PhDs of the CRETE\_CSD department.}
\label{tab-top-10-PhDs-CRETE-CSD}
\end{table}

\begin{table}[!hbt]
\center
\begin{tabular}{||c|c|c||}\hline\hline
{\bf {\#}articles}                  & {\bf {\#}citations}                                                           & $h$-{\bf index}     \\\hline
{\scriptsize Kosmatopoulos ($125$)} & {\scriptsize Koutroulis ($2809$)}                                             & {\scriptsize Kolokotsa ($26$)} \\\hline
{\scriptsize Sakkalis ($110$)}      & {\scriptsize Kosmatopoulos ($2432$)}                                          & {\scriptsize Kosmatopoulos ($25$)} \\\hline
{\scriptsize Rovithakis ($103$)}    & {\scriptsize Kolokotsa ($2263$)}                                              & \minitab[c]{{\scriptsize Koutroulis ($20$)}\\{\scriptsize Rovithakis ($20$)}} \\\hline
{\scriptsize Koutsakis ($101$)}     & {\scriptsize Rovithakis ($1788$)}                                             & {\scriptsize Dounis ($18$)} \\\hline
{\scriptsize Kolokotsa ($84$)}      & {\scriptsize Dounis ($1102$)}                                                 & {\scriptsize Sakkalis ($14$)} \\\hline
{\scriptsize Kornaros ($59$)}       & {\scriptsize Sakkalis ($946$)}                                                & {\scriptsize Karipidis ($13$)} \\\hline
{\scriptsize Koutroulis ($56$)}     & {\scriptsize Karipidis ($672$)}                                               & {\scriptsize Koutsakis ($11$)} \\\hline
{\scriptsize Sfakianakis ($52$)}    & {\scriptsize Koutsakis ($345$)}                                               & {\scriptsize Tsinaraki ($10$)} \\\hline
{\scriptsize Dounis ($47$)}         & \minitab[c]{{\scriptsize Manolakis ($322$)}\\{\scriptsize Tsinaraki ($322$)}} & \multirow{2}*{}\\\cline{1-2}
{\scriptsize Tryfonopoulos ($40$)}  & {\scriptsize Raftopoulou ($282$)}                                             & \\\cline{1-2}\hline\hline
\end{tabular}
\vspace*{.05\baselineskip}
\caption{Top-$10$ PhDs of the CRETE\_ECE department.}
\label{tab-top-10-PhDs-CRETE-ECE}
\end{table}

\begin{table}[!hbt]
\center
\begin{tabular}{||c|c|c||}\hline\hline
{\bf {\#}articles}                                                              & {\bf {\#}citations}                                                      & $h$-{\bf index}     \\\hline
{\scriptsize Tsipouras ($72$)}                                                  & {\scriptsize Tsipouras ($1186$)}                                         & {\scriptsize Tsipouras ($17$)} \\\hline
{\scriptsize Rigas ($61$)}                                                      & {\scriptsize Tsoulos ($481$)}                                            & {\scriptsize Tsoulos ($14$)} \\\hline
\minitab[c]{{\scriptsize Papadopoulos ($46$)}\\{\scriptsize Stefanidis ($46$)}} & {\scriptsize Stefanidis ($405$)}                                         & {\scriptsize Rigas ($13$)} \\\hline
{\scriptsize Tripoliti ($44$)}                                                  & {\scriptsize Rigas ($349$)}                                              & {\scriptsize Stefanidis ($11$)} \\\hline
{\scriptsize Tsoulos ($43$)}                                                    & {\scriptsize Plissiti ($343$)}                                           & \multirow{6}*{} \\\cline{1-2}
{\scriptsize Karvelis ($39$)}                                                   & \minitab[c]{{\scriptsize Tzikas ($294$)}\\{\scriptsize Chantas ($294$)}} & \\\cline{1-2}
{\scriptsize Tenentes ($24$)}                                                   & {\scriptsize Drosou ($243$)}                                             & \\\cline{1-2}
{\scriptsize Voglis ($23$)}                                                     & {\scriptsize Constantinopoulos ($242$)}                                  & \\\cline{1-2}
{\scriptsize Koloniari ($22$)}                                                  & {\scriptsize Papadopoulos ($235$)}                                       & \\\cline{1-2}
{\scriptsize Sfikas ($20$)}                                                     & {\scriptsize Tripoliti ($211$)}                                          & \\\cline{1-2}\hline\hline
\end{tabular}
\vspace*{.05\baselineskip}
\caption{Top-$10$ PhDs of the IOAN\_CSE department.}
\label{tab-top-10-PhDs-IOAN-CSE}
\end{table}

\begin{table}[!hbt]
\center
\begin{tabular}{||c|c|c||}\hline\hline
{\bf {\#}articles}                                                                              & {\bf {\#}citations}                                                   & $h$-{\bf index}     \\\hline
{\scriptsize Giannakos ($104$)}                                                                 & {\scriptsize Giannakos ($444$)}                                       & {\scriptsize Giannakos ($10$)} \\\hline
{\scriptsize Alexiou ($21$)}                                                                    & {\scriptsize Mikalef ($59$)}                                          & \multirow{8}*{} \\\cline{1-2}
{\scriptsize Mikalef ($20$)}                                                                    & {\scriptsize Pappas ($53$)}                                           & \\\cline{1-2}
{\scriptsize Pappas ($19$)}                                                                     & {\scriptsize Alexiou ($40$)}                                          & \\\cline{1-2}
{\scriptsize Ringas ($10$)}                                                                     & {\scriptsize Ringas ($31$)}                                           & \\\cline{1-2}
{\scriptsize Giannakis ($9$)}                                                                   & {\scriptsize Lapatas ($7$)}                                           & \\\cline{1-2}
{\scriptsize Panaretos ($6$)}                                                                   & {\scriptsize Kontzalis ($6$)}                                         & \\\cline{1-2}
\minitab[c]{{\scriptsize Lapatas ($5$)}\\{\scriptsize Plerou ($5$)}\\{\scriptsize Psiha ($5$)}} & \minitab[c]{{\scriptsize Giannakis ($5$)}\\{\scriptsize Psiha ($5$)}} & \\\cline{1-2}
{\scriptsize Kontzalis ($3$)}                                                                   & {\scriptsize Panaretos ($4$)}                                         & \\\cline{1-2}
{\scriptsize NO OTHER PhDs}                                                                     & {\scriptsize Plerou ($2$)}                                            & \\\cline{1-2}\hline\hline
\end{tabular}
\vspace*{.05\baselineskip}
\caption{Top-$10$ PhDs of the IONIO\_DI department.}
\label{tab-top-10-PhDs-IONIO-DI}
\end{table}

\begin{table}[!hbt]
\center
\begin{tabular}{||@{}c@{}|@{}c@{}|@{}c@{}||}\hline\hline
{\bf {\#}articles}                    & {\bf {\#}citations}                  & $h$-{\bf index}     \\\hline
{\scriptsize Demestichas, P. ($262$)} & {\scriptsize Papadias ($7099$)}      & {\scriptsize Papadias ($44$)} \\\hline
{\scriptsize Panagopoulos A. ($256$)} & {\scriptsize Efremidis ($4384$)}     & {\scriptsize Vlassis ($28$)} \\\hline
{\scriptsize Maglogiannis ($252$)}    & {\scriptsize Vlassis ($3201$)}       & \minitab[c]{{\scriptsize Efremidis ($26$)}\\{\scriptsize Papathanassiou ($26$)}} \\\hline
{\scriptsize Doulamis, A. ($224$)}    & {\scriptsize Papathanasiou ($2853$)} & {\scriptsize Simitsis ($25$)} \\\hline
{\scriptsize Pleros ($208$)}          & {\scriptsize Dimeas ($2799$)}        & {\scriptsize Vassiliadis, Pan. ($24$)} \\\hline
{\scriptsize Rigatos ($199$)}         & {\scriptsize Kontopantelis ($2674$)} & \minitab[c]{{\scriptsize Avrithis ($23$)}\\{\scriptsize Georgilakis ($23$)}\\{\scriptsize Maglogiannis ($23$)}\\{\scriptsize Pleros ($23$)}}\\\hline
{\scriptsize Orfanelli ($178$)}       & {\scriptsize Georgilakis ($1996$)}   & \minitab[c]{{\scriptsize Avgeriou ($21$)}\\{\scriptsize Demestichas, P. ($21$)}\\{\scriptsize Rigatos ($21$)}}\\\hline
{\scriptsize Doulamis, N. ($172$)}    & {\scriptsize Dalamagas ($1857$)}     & \minitab[c]{{\scriptsize Doulamis, A. ($20$)}\\{\scriptsize Doulamis, N. ($20$)}\\{\scriptsize Vlachos ($20$)}\\{\scriptsize Zoiros ($20$)}}\\\hline
{\scriptsize Avgeriou ($170$)}        & {\scriptsize Koziris ($1783$)}       & \minitab[c]{{\scriptsize Doukas ($19$)}\\{\scriptsize Kokkinos ($19$)}\\{\scriptsize Koziris ($19$)}}\\\hline
{\scriptsize Sygletos ($167$)}        & {\scriptsize Maglogiannis ($1777$)}  & \minitab[c]{{\scriptsize Anagnostopoulos ($18$)}\\{\scriptsize Igoumenidis ($18$)}\\{\scriptsize Paliatsos ($18$)}\\{\scriptsize Panagopoulos A. ($18$)}\\{\scriptsize Peppas ($18$)}}\\\hline
\end{tabular}
\vspace*{.05\baselineskip}
\caption{Top-$10$ PhDs of the NTUA\_ECE department.}
\label{tab-top-10-PhDs-NTUA-ECE}
\end{table}

\begin{table}[!hbt]
\center
\begin{tabular}{||@{}c@{}|@{}c@{}|@{}c@{}||}\hline\hline
{\bf {\#}articles}                                                                    & {\bf {\#}citations}                   & $h$-{\bf index}     \\\hline
{\scriptsize Soudris ($367$)}                                                         & {\scriptsize Karagiannidis ($6561$)}  & {\scriptsize Karagiannidis ($43$)} \\\hline
{\scriptsize Karagiannidis ($354$)}                                                   & {\scriptsize Andrikopoulos ($2539$)}  & {\scriptsize Papageorgiou ($25$)} \\\hline
{\scriptsize Fakotakis ($200$)}                                                       & {\scriptsize Tsiftsis ($2455$)}       & \minitab[c]{{\scriptsize Stylios ($24$)}\\{\scriptsize Tsiftsis ($24$)}} \\\hline
{\scriptsize Magoulas ($159$)}                                                        & {\scriptsize Papageorgiou ($2335$)}   & \minitab[c]{{\scriptsize Andrikopoulos ($22$)}\\{\scriptsize Magoulas ($22$)}} \\\hline
{\scriptsize Koufopavlou ($151$)}                                                     & {\scriptsize Stylios ($2245$)}        & \minitab[c]{{\scriptsize Fakotakis  ($19$)}\\{\scriptsize Stamatatos ($19$)}} \\\hline
{\scriptsize Andrikopoulos ($150$)}                                                   & {\scriptsize Magoulas ($2015$)}       & \minitab[c]{{\scriptsize Koufopavlou ($17$)}\\{\scriptsize Nikolakopoulos ($17$)}} \\\hline
{\scriptsize Nikolaidis ($144$)}                                                      & {\scriptsize Stamatatos ($1468$)}     & {\scriptsize Sklavos ($16$)} \\\hline
\minitab[c]{{\scriptsize Nikolakopoulos ($138$)}\\{\scriptsize Papageorgiou ($138$)}} & {\scriptsize Fakotakis ($1434$)}      & \minitab[c]{{\scriptsize Alexis ($15$)}\\{\scriptsize Kalis ($15$)}} \\\hline
{\scriptsize Stylios ($136$)}                                                         & {\scriptsize Antonakopoulos ($1230$)} & \minitab[c]{{\scriptsize Soudris ($14$)}\\{\scriptsize Zogas ($14$)}} \\\hline
{\scriptsize Paliouras ($109$)}                                                       & {\scriptsize Koufopavlou ($1111$)}    & \minitab[c]{{\scriptsize Ganchev ($13$)}\\{\scriptsize Logothetis ($13$)}\\{\scriptsize Potamitis ($13$)}} \\\hline
\end{tabular}
\vspace*{.05\baselineskip}
\caption{Top-$10$ PhDs of the PATRAS\_ECE department.}
\label{tab-top-10-PhDs-PATRAS-ECE}
\end{table}

\begin{table}[!hbt]
\center
\begin{tabular}{||c|c|c||}\hline\hline
{\bf {\#}articles}                & {\bf {\#}citations}                 & $h$-{\bf index}     \\\hline
{\scriptsize Monovasilis ($60$)}  & {\scriptsize Giannopoulou ($1638$)} & {\scriptsize Anastassi ($21$)} \\\hline
{\scriptsize Anastassi ($58$)}    & {\scriptsize Anastassi ($1164$)}    & \minitab[c]{{\scriptsize Giannopoulou ($16$)}\\{\scriptsize Monovasilis ($16$)}} \\\hline
{\scriptsize Papadopoulos ($40$)} & {\scriptsize Monovasilis ($906$)}   & {\scriptsize Papadopoulos ($12$)} \\\hline
{\scriptsize Uzunidis ($35$)}     & {\scriptsize Papadopoulos ($813$)}  & {\scriptsize Panopoulos ($10$)} \\\hline
{\scriptsize Giannopoulou ($34$)} & {\scriptsize Sakas ($233$)}         &  \multirow{6}*{}\\\cline{1-2}
{\scriptsize Rizos ($33$)}        & {\scriptsize Panopoulos ($186$)}    &  \\\cline{1-2}
{\scriptsize Vasiliadis ($32$)}   & {\scriptsize Kosti ($174$)}         &  \\\cline{1-2}
{\scriptsize Sakas ($29$)}        & {\scriptsize Tselios ($137$)}       &  \\\cline{1-2}
{\scriptsize Antoniou ($21$)}     & {\scriptsize Rizos ($77$)}          &  \\\cline{1-2}
{\scriptsize Kosmas ($18$)}       & {\scriptsize Vasiliadis ($76$)}     &  \\\cline{1-2}\hline\hline
\end{tabular}
\vspace*{.05\baselineskip}
\caption{Top-$10$ PhDs of the PELOP\_DI department.}
\label{tab-top-10-PhDs-PELOP-DI}
\end{table}

\begin{table}[!hbt]
\center
\begin{tabular}{||@{}c@{}|@{}c@{}|@{}c@{}||}\hline\hline
{\bf {\#}articles}                                                                     & {\bf {\#}citations}                     & $h$-{\bf index}     \\\hline
{\scriptsize Bogris ($166$)}                                                           & {\scriptsize Bogris ($2057$)}           & {\scriptsize Bogris ($24$)} \\\hline
{\scriptsize Anagnostopoulos ($85$)}                                                   & {\scriptsize Argyris ($1578$)}          & {\scriptsize Koutrika ($19$)} \\\hline
{\scriptsize Koutrika ($75$)}                                                          & {\scriptsize Koutrika ($1492$)}         & {\scriptsize Argyris ($17$)} \\\hline
{\scriptsize Argyris ($71$)}                                                           & {\scriptsize Passalis ($832$)}          & \minitab[c]{{\scriptsize Christodoulou ($14$)}\\{\scriptsize Passalis ($14$)}\\{\scriptsize Stamatopoulos ($14$)}} \\\hline
{\scriptsize Kapsalis ($68$)}                                                          & {\scriptsize Christodoulou ($742$)}     & {\scriptsize Tsigaridas ($13$)} \\\hline
{\scriptsize Simos ($59$)}                                                             & {\scriptsize Papadimitropoulos ($610$)} & \minitab[c]{{\scriptsize Anagnostopoulos ($12$)}\\{\scriptsize Kranitis ($12$)}\\{\scriptsize Louloudis ($12$)}\\{\scriptsize Papadimitropoulos ($12$)}\\{\scriptsize Simos ($12$)}} \\\hline
\minitab[@{}c@{}]{{\scriptsize Bouboulis ($57$)}\\{\scriptsize Giannakopoulos ($57$)}} & {\scriptsize Anagnostopoulos ($579$)}   & {\scriptsize Ntirogiannis ($11$)} \\\hline
{\scriptsize Koumaras ($51$)}                                                          & {\scriptsize Ntirogiannis ($532$)}      & \minitab[c]{{\scriptsize Bouboulis ($10$)}\\{\scriptsize Giannakopoulos ($10$)}\\{\scriptsize Mavroforakis ($10$)}\\{\scriptsize Papadakis ($10$)}} \\\hline
{\scriptsize Tsigaridas ($50$)}                                                        & {\scriptsize Stamatopoulos ($529$)}     & \multirow{2}*{}\\\cline{1-2}
{\scriptsize Cutsuridis ($49$)}                                                        & {\scriptsize Papadakis ($500$)}         & \\\cline{1-2}\hline\hline
\end{tabular}
\vspace*{.05\baselineskip}
\caption{Top-$10$ PhDs of the UA\_DI department.}
\label{tab-top-10-PhDs-UA-DI}
\end{table}

\begin{table}[!hbt]
\center
\begin{tabular}{||@{}c|c|c@{}||}\hline\hline
{\bf {\#}articles}                                                             & {\bf {\#}citations}                 & $h$-{\bf index}     \\\hline
{\scriptsize Alepis ($70$)}                                                    & {\scriptsize Mitrokosta ($704$)}    & {\scriptsize Mitrokosta ($14$)} \\\hline
\minitab[c]{{\scriptsize Mitrokosta ($63$)}\\{\scriptsize Patsakis ($63$)}}    & {\scriptsize Kotzanikolaou ($469$)} & {\scriptsize Kotzanikolaou ($12$)} \\\hline
{\scriptsize Kambasi ($60$)}                                                   & {\scriptsize Frentzos ($456$)}      & {\scriptsize Frentzos ($11$)} \\\hline
{\scriptsize Kotzanikolaou ($53$)}                                             & {\scriptsize Poulos ($406$)}        & {\scriptsize Ntoutsi ($10$)} \\\hline
{\scriptsize Poulos ($50$)}                                                    & {\scriptsize Ntoutsi ($394$)}       & \multirow{6}*{} \\\cline{1-2}
\minitab[c]{{\scriptsize Lambropoulos ($40$)}\\{\scriptsize Magkos ($40$)}}    & {\scriptsize Nikolidakis ($391$)}   & \\\cline{1-2}
{\scriptsize Vosinakis ($38$)}                                                 & {\scriptsize Katsionis ($331$)}     & \\\cline{1-2}
\minitab[c]{{\scriptsize Papadakis ($32$)}\\{\scriptsize Stathopoulou ($32$)}} & {\scriptsize Alepis ($289$)}        & \\\cline{1-2}
{\scriptsize Sotiropoulos ($27$)}                                              & {\scriptsize Patsakis ($276$)}      & \\\cline{1-2} 
{\scriptsize Ntoutsi ($26$)}                                                   & {\scriptsize Giatrakos ($274$)}     & \\\cline{1-2}\hline\hline
\end{tabular}
\vspace*{.05\baselineskip}
\caption{Top-$10$ PhDs of the UNIPI\_DI department.}
\label{tab-top-10-PhDs-UNIPI-DI}
\end{table}

\begin{table}[!hbt]
\center
\begin{tabular}{||c|c|c||}\hline\hline
{\bf {\#}articles}                                                          & {\bf {\#}citations}                 & $h$-{\bf index}     \\\hline
{\scriptsize Tasoulis ($25$)}                                               & {\scriptsize Dimou ($206$)}         & {\scriptsize Dimou ($5$)} \\\hline
{\scriptsize Xanthis ($16$)}                                                & {\scriptsize Tasoulis ($63$)}       & \multirow{5}*{}\\\cline{1-2}
{\scriptsize Moutselos ($13$)}                                              & {\scriptsize Moutselos ($62$)}      & \\\cline{1-2}
{\scriptsize Dimou ($11$)}                                                  & {\scriptsize Xanthis ($32$)}        & \\\cline{1-2}
\minitab[c]{{\scriptsize Haralabopoulos ($4$)}\\{\scriptsize Kontou ($4$)}} & {\scriptsize Haralabopoulos ($11$)} & \\\cline{1-2}
{\scriptsize NO OTHER PhDs}                                                 & {\scriptsize Kontou ($5$)}          & \\\cline{1-2}\hline\hline
\end{tabular}
\vspace*{.05\baselineskip}
\caption{Top-$10$ PhDs of the UTH\_DIB department.}
\label{tab-top-10-PhDs-UTH-DIB}
\end{table}

\begin{table}[!hbt]
\center
\begin{tabular}{||@{}c|c|c@{}||}\hline\hline
{\bf {\#}articles}                                                            & {\bf {\#}citations}                     & $h$-{\bf index}     \\\hline
{\scriptsize Korakis ($139$)}                                                 & {\scriptsize Korakis ($1419$)}          & {\scriptsize Korakis ($17$)} \\\hline
{\scriptsize Gkoulalas-Divanis ($65$)}                                        & {\scriptsize Georgakilas ($1096$)}      & {\scriptsize Gkoulalas-Divanis ($14$)} \\\hline
{\scriptsize Iosifidis ($47$)}                                                & {\scriptsize Paraskevopoulou ($818$)}   & {\scriptsize Iosifidis ($13$)} \\\hline
{\scriptsize Sourlas ($43$)}                                                  & {\scriptsize Gkoulalas-Divanis ($663$)} & {\scriptsize Axenopoulos ($12$)} \\\hline
\minitab[c]{{\scriptsize Maglaras ($35$)}\\{\scriptsize Tziritas ($35$)}}     & {\scriptsize Iosifidis ($477$)}         & {\scriptsize Sourlas ($11$)} \\\hline
{\scriptsize Athanasiou ($32$)}                                               & {\scriptsize Axenopoulos ($404$)}       & \multirow{5}*{} \\\cline{1-2}
\minitab[c]{{\scriptsize Axenopoulos ($30$)}\\{\scriptsize Syrivelis ($30$)}} & {\scriptsize Sourlas ($350$)}           & \\\cline{1-2}
{\scriptsize Katsalis ($23$)}                                                 & {\scriptsize Gkatzikis ($250$)}         & \\\cline{1-2}
\minitab[c]{{\scriptsize Gkatzikis ($22$)}\\{\scriptsize Poularakis ($22$)}}  & {\scriptsize Athanasiou ($187$)}        & \\\cline{1-2} 
\minitab[c]{{\scriptsize Keranidis ($20$)}\\{\scriptsize Choumas ($20$)}}     & {\scriptsize Poularakis ($183$)}        & \\\cline{1-2}\hline\hline
\end{tabular}
\vspace*{.05\baselineskip}
\caption{Top-$10$ PhDs of the UTH\_ECE department.}
\label{tab-top-10-PhDs-UTH-ECE}
\end{table}

Finally, Table~\ref{top-50-PhDs-ranked-by-impact} presents the top-$50$ PhDs based on impact. As expected we will recognize many current faculty members
in national or international universities being in the top positions of this list.

\begin{table}[!hbt]
\center
\begin{tabular}{||l|l|c||}\hline\hline
1	&	Papadias            &	NTUA\_ECE	\\\hline
2	&	Karagiannidis       &	PATRAS\_ECE	\\\hline
3	&	Efremidis           &	NTUA\_ECE	\\\hline
4	&	Vlassis             &	NTUA\_ECE	\\\hline
5	&	Papathanasiou Stav. &	NTUA\_ECE	\\\hline
6	&	Koutroulis          &	CRETE\_ECE	\\\hline
7	&	Dimeas              &	NTUA\_ECE	\\\hline
8	&	Tsoumakas           & 	AUTH\_DI	\\\hline
9	&	Kontopantelis       &	NTUA\_ECE	\\\hline
10	&	Andirkopoulos       &	PATRAS\_ECE	\\\hline
11	&	Tsiftsis            &	PATRAS\_ECE	\\\hline
12	&	Kosmatopoulos       &	CRETE\_ECE	\\\hline
13	&	Zafeiriou           &	AUTH\_DI	\\\hline
14	&	Tefas               &	AUTH\_DI	\\\hline
15	&	Papageorgiou Elp.   &	PATRAS\_ECE	\\\hline
16	&	Kolokotsa           &	CRETE\_ECE	\\\hline
17	&	Stylios             &	PATRAS\_ECE	\\\hline
18	&	Komodakis           &	CRETE\_CSD	\\\hline
19	&	Halkidi             &	AUEB\_DI	\\\hline
20	&	Argyros             &	CRETE\_CSD	\\\hline
21	&	Nanopoulos          &	AUTH\_DI	\\\hline
22	&	Bogris              &	UA\_DI	\\\hline
23	&	Magoulas            &	PATRAS\_ECE	\\\hline
24	&	Georgilakis         &	NTUA\_ECE	\\\hline
25	&	Vakali              &	AUTH\_DI	\\\hline
26	&	Dalamagas           &	NTUA\_ECE	\\\hline
27	&	Rovithakis          &	CRETE\_ECE	\\\hline
28	&	Koziris             &	NTUA\_ECE	\\\hline
29	&	Maglogiannis        &	NTUA\_ECE	\\\hline
30	&	Katsaros D.         &	AUTH\_DI	\\\hline
31	&	Vassiliadis         &	NTUA\_ECE	\\\hline
32	&	Simitsis            &	NTUA\_ECE	\\\hline
33	&	Vergoulis           &	NTUA\_ECE	\\\hline
34	&	Pleros              &	NTUA\_ECE	\\\hline
35	&	Avgeriou            &	NTUA\_ECE	\\\hline
36	&	Giannopoulou Evg.   &	PEL\_DI	\\\hline
37	&	Demestichas         &	NTUA\_ECE	\\\hline
38	&	Petrakis            &	CRETE\_CSD	\\\hline
39	&	Argyris             &	UA\_DI	\\\hline
40	&	Tsapatsoulis        &	NTUA\_ECE	\\\hline
41	&	Panagopoulos        &	NTUA\_ECE	\\\hline
42	&	Koutrika            &	UA\_DI	\\\hline
43	&	Stamatatos          &	PATRAS\_ECE	\\\hline
44	&	Avrithis            &	NTUA\_ECE	\\\hline
45	&	Michalopoulos       &	AUTH\_ECE	\\\hline
46	&	Fakotakis           &	PATRAS\_ECE	\\\hline
47	&	Siris               &	CRETE\_CSD	\\\hline
48	&	Korakis             &	UTH\_ECE	\\\hline
49	&	Doulamis A.         &	NTUA\_ECE	\\\hline
50	&	Mezaris             &	AUTH\_ECE	\\\hline\hline
\end{tabular}
\vspace*{.05\baselineskip}
\caption{Top-$50$ PhDs across all departments according to the total number of citations.}
\label{top-50-PhDs-ranked-by-impact}
\end{table}

\begin{figure}[!hbt]
\begin{center}
\includegraphics[scale=.7]{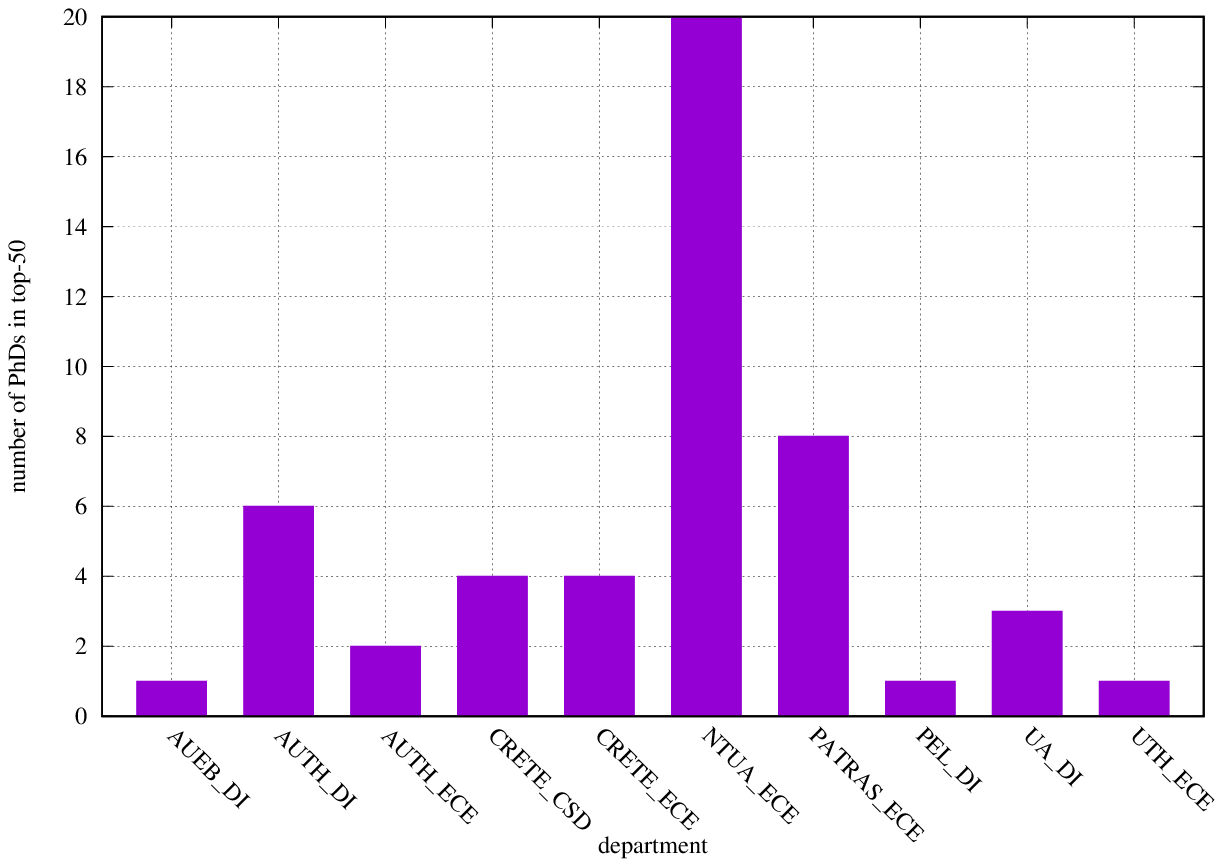}
\end{center}
\vspace*{-\baselineskip}
\caption{Departments' contributions in top-$50$ impactful PhDs.}
\label{fig-dept-contrib-top-50}
\end{figure}

Figure~\ref{fig-dept-contrib-top-50} presents the contribution of each department in populating the list of top-$50$, where NTUA\_ECE is the dominant department,
and CRETE\_ECE has significant presence even though has not awarded many PhDs.

\section{Rankings of departments}
\label{sec-depts-rankings}

There are several hundreds of bibliometric indicators~\cite{Todeschini-Handbook16} to describe different aspects of the research productivity and impact of a 
scientist, and then based on individuals' performance to describe organizations' performance, e.g., of a university. In this work, we have used the most common 
such measures, namely, the number of published articles as the productivity measure, the number of citations as the impact measure, and the 
$h$-index~\cite{Hirsch-PNAS05} as a proxy for both productivity and impact.

For each department, we have calculated the median and average values for articles, citations, and $h$-index based on the respective PhDs' data. We emphasize at 
this point two issues: a) we recommend the {\it median as the most appropriate performance measure}, and b) when there are a {\it lot of ties, then we consider 
the average value as the most credible performance measure}. The reason is due to the nature of the underlying distributions of the number of published articles 
and received citations. These distributions are high-skewed, i.e., power-law, and a few individuals are responsible for the vast majority of the performance. 
For such distributions, the mean is not a representative measure; median is preferable. On the other hand, in cases with many ties, e.g., the $h$-index case, 
then the average values can be used for comparison purposes.

The results are presented in Tables~\ref{tab-dept-ranking-by-productivity}--\ref{tab-dept-ranking-by-h-index}. The most strong and obvious result is that there 
is a `winner' department, namely CRETE\_ECE, and the other two positions in top-$3$ are occupied by AUTH\_DI and CRETE\_CSD. We were not surprised to see the
`Cretan' departments in the top positions, even though they are `peripherals'. We had established the research excellence of CRETE\_CSD's faculty in our past 
work~\cite{Katsaros-PCI08}, and here we confirm their steady dedication to breathe their excellence into their PhD students.

Focusing on median citation-based impact, which we consider a rather {\it stable and very significant} performance measure, it is promising to see `peripheral' 
departments, namely AEGEAN\_ICSD and IOAN\_CSE to be located immediately after the top-$3$ and to overcome almost all (except from AUTH\_DI) `central' departments.

The `large', old engineering departments, namely AUTH\_ECE, NTUA\_ECE and PATRAS\_ECE have moderate performance both in terms of productivity and impact, and are
located at the middle of the ranking lists when examining their median values. Their performance is slightly better with respect to the average values. An 
interesting observation is that they perform rather bad with respect to the average $h$-index, which can be explained by the performance of their individual 
PhDs, i.e., very few of them have quite large $h$-index, and the vast majority of PhDs has a single-digit $h$-index value.

\begin{table}[!hbt]
\center
\begin{tabular}{||@{}l|@{}c@{}|||@{}l|@{}c@{}||}\hline\hline
{\bf Department}           & {\bf median {\#}articles} & {\bf Department}           & {\bf avg {\#}articles}\\\hline
{\scriptsize CRETE\_ECE}   & $20$                      & {\scriptsize AUTH\_DI}     & $32.6$\\\hline
{\scriptsize AUTH\_DI}     & $18$                      & {\scriptsize CRETE\_ECE}   & $31.1$\\\hline
{\scriptsize CRETE\_CSD}   & $18$                      & {\scriptsize CRETE\_CSD}   & $28.6$\\\hline
{\scriptsize UTH\_ECE}     & $15$                      & {\scriptsize PATRAS\_ECE}  & $25.0$\\\hline
{\scriptsize AEGEAN\_ICSD} & $14$                      & {\scriptsize NTUA\_ECE}    & $21.7$\\\hline
{\scriptsize PELOP\_DI}    & $14$                      & {\scriptsize AEGEAN\_ICSD} & $20.6$\\\hline
{\scriptsize AUTH\_ECE}    & $13$                      & {\scriptsize PELOP\_DI}    & $19.7$\\\hline
{\scriptsize IOAN\_CSE}    & $13$                      & {\scriptsize UTH\_ECE}     & $19.7$\\\hline
{\scriptsize PATRAS\_ECE}  & $13$                      & {\scriptsize AUTH\_ECE}    & $19.4$\\\hline
{\scriptsize UTH\_DIB}     & $12$                      & {\scriptsize IONIO\_DI}    & $18.8$\\\hline
{\scriptsize NTUA\_ECE}    & $11$                      & {\scriptsize IOAN\_CSE}    & $18.0$\\\hline
{\scriptsize UA\_DI}       & $11$                      & {\scriptsize UA\_DI}       & $17.4$\\\hline
{\scriptsize AUEB\_DI}     & $10$                      & {\scriptsize AUEB\_DI}     & $17.2$\\\hline
{\scriptsize IONIO\_DI}    &  $9$                      & {\scriptsize UNIPI\_DI}    & $12.8$\\\hline
{\scriptsize UNIPI\_DI}    &  $7$                      & {\scriptsize UTH\_DIB}     & $12.2$\\\hline\hline
\end{tabular}
\vspace*{.05\baselineskip}
\caption{Ranking departments by median and average productivity.}
\label{tab-dept-ranking-by-productivity}
\end{table}

\begin{table}[!hbt]
\center
\begin{tabular}{||@{}l|@{}c@{}|||@{}l|@{}c@{}||}\hline\hline
{\bf Department}           & {\bf median {\#}citations}   & {\bf Department}           & {\bf avg {\#}citations}\\\hline
{\scriptsize CRETE\_ECE}   & $125$                        & {\scriptsize CRETE\_ECE}   & $380.9$\\\hline
{\scriptsize AUTH\_DI}     & $125$                        & {\scriptsize AUTH\_DI}     & $312.7$\\\hline
{\scriptsize CRETE\_CSD}   & $120$                        & {\scriptsize CRETE\_CSD}   & $292.3$\\\hline
{\scriptsize AEGEAN\_ICSD} & $79$                         & {\scriptsize PELOP\_DI}    & $250.7$\\\hline
{\scriptsize IOAN\_CSE}    & $73$                         & {\scriptsize PATRAS\_ECE}  & $186.0$\\\hline
{\scriptsize UA\_DI}       & $70$                         & {\scriptsize NTUA\_ECE}    & $178.4$\\\hline
{\scriptsize AUTH\_ECE}    & $63$                         & {\scriptsize AUEB\_DI}     & $173.4$\\\hline
{\scriptsize PATRAS\_ECE}  & $63$                         & {\scriptsize UTH\_ECE}     & $171.2$\\\hline
{\scriptsize NTUA\_ECE}    & $54$                         & {\scriptsize IOAN\_CSE}    & $151.0$\\\hline
{\scriptsize PELOP\_DI}    & $53$                         & {\scriptsize AUTH\_ECE}    & $150.0$\\\hline
{\scriptsize UTH\_ECE}     & $50$                         & {\scriptsize UA\_DI}       & $146.5$\\\hline
{\scriptsize AUEB\_DI}     & $47$                         & {\scriptsize AEGEAN\_ICSD} & $131.2$\\\hline
{\scriptsize UTH\_DIB}     & $47$                         & {\scriptsize UNIPI\_DI}    & $83.7$\\\hline
{\scriptsize UNIPI\_DI}    & $19$                         & {\scriptsize UTH\_DIB}     & $63.2$\\\hline
{\scriptsize IONIO\_DI}    & $7$                          & {\scriptsize IONIO\_DI}    & $59.6$\\\hline\hline
\end{tabular}
\vspace*{.05\baselineskip}
\caption{Ranking departments by median and average impact.}
\label{tab-dept-ranking-by-impact}
\end{table}

\begin{table}[!hbt]
\center
\begin{tabular}{||@{}l|@{}c@{}|||@{}l|@{}c@{}||}\hline\hline
{\bf Department}           & {\bf median $h$-index}   & {\bf Department}           & {\bf avg $h$-index}\\\hline
{\scriptsize CRETE\_ECE}   & $6$                      & {\scriptsize CRETE\_ECE}   & $7.5$\\\hline
{\scriptsize AUTH\_DI}     & $5$                      & {\scriptsize AUTH\_DI}     & $6.9$\\\hline
{\scriptsize CRETE\_CSD}   & $5$                      & {\scriptsize CRETE\_CSD}   & $6.7$\\\hline
{\scriptsize AEGEAN\_ICSD} & $4$                      & {\scriptsize PELOP\_DI}    & $5.7$\\\hline
{\scriptsize AUEB\_DI}     & $4$                      & {\scriptsize AEGEAN\_ICSD} & $5.3$\\\hline
{\scriptsize AUTH\_ECE}    & $4$                      & {\scriptsize IOAN\_CSE}    & $5.3$\\\hline
{\scriptsize IOAN\_CSE}    & $4$                      & {\scriptsize AUEB\_DI}     & $5.0$\\\hline
{\scriptsize NTUA\_ECE}    & $4$                      & {\scriptsize AUTH\_ECE}    & $5.0$\\\hline
{\scriptsize PATRAS\_ECE}  & $4$                      & {\scriptsize PATRAS\_ECE}  & $5.0$\\\hline
{\scriptsize PELOP\_DI}    & $4$                      & {\scriptsize UA\_DI}       & $5.0$\\\hline
{\scriptsize UA\_DI}       & $4$                      & {\scriptsize NTUA\_ECE}    & $4.9$\\\hline
{\scriptsize UTH\_DIB}     & $4$                      & {\scriptsize UTH\_ECE}     & $4.8$\\\hline
{\scriptsize UTH\_ECE}     & $4$                      & {\scriptsize UTH\_DIB}     & $3.5$\\\hline
{\scriptsize UNIPI\_DI}    & $3$                      & {\scriptsize UNIPI\_DI}    & $3.4$\\\hline
{\scriptsize IONIO\_DI}    & $2$                      & {\scriptsize IONIO\_DI}    & $3.2$\\\hline\hline
\end{tabular}
\vspace*{.05\baselineskip}
\caption{Ranking departments by median and average $h$-index.}
\label{tab-dept-ranking-by-h-index}
\end{table}

\section{Data collection}
\label{sec-data-collection}

It is worth mentioning that very few departments maintain publicly accessible {\it electronic} files for their PhD graduates, namely 
AEGEAN\_ICSD~\cite{AEGEANICSDI-PhDs}, CRETE\_CSD~\cite{CRETECSD-PhDs}, UA\_DI~\cite{UADI-PhDs}, and UTH\_ECE~\cite{UTHECE-PhDs}. Moreover, some departments do 
not maintain {\it electronic} files at all for their graduates. Thus, we asked for data -- via our colleagues -- from the administrative staff of the departments. 
Unfortunately, not all administrations were able to provide assistance.

We decided to gather the performance data from Scopus instead of Google Scholar or ISI Web of Science. There exist rich literature comparing these bibliographic 
databases, see for instance~\cite{Bar-Ilan-Scientometrics08}. Our main criterion was that Scopus database is cleansed, and it contains
both journal and conference publications, which is significant to our study taking into account that conference publications are considered equally important in 
computer science~\cite{Meyer-CACM09,Vrettas-JASIST15} compared to other disciplines. Especially for Google Scholar, analysis of 
literature~\cite{Halevi-JNL-Informetrics17} concludes that it currenlty lacks quality control, clear indexing guidelines, and it can be easily manipulated.

We collected our data mostly manually (by browsing and searching). Scopus provides an API for retrieving data, but the offered data are not as 
up-to-date\footnote{The following information is undocumented. There are two different databases that Scopus uses, one for the Web, and one for the API, and 
they do not return the same data. It is close, but not the same; the API database may not be updated as quickly as the Web database.} as those returned to user 
browsing. Manual data collection turned out to be the fastest and most effective method. Should we have used Scopus `Search Form' to search for a specific PhD, 
then we would have very frequently faced the case, where the Scopus maintains different profiles for persons with exactly the same name, making the disambiguation 
more time-consuming. Thus, we started our browsing from a supervisor's profile which was relatively easy to discover, because of the wealth of publications s/he 
has accumulated over the years. Then, we sought for the names of his/her PhDs. This methodology allowed us to detect the correct PhD's profile in the vast 
majority of the cases. A first problem arose with the existence of what we call the `mixed profiles'; a mixed profile is a single Scopus profile which contains 
articles belonging to different persons -- who share exactly the same name though. We had to manually and very carefully cleanse such profiles (performing joins 
with other bibliographic databases), because their impact on the final data could be dramatic. For instance, NTUA\_ECE's PhD Ioannis 
Konstantinou\footnote{\tiny https://www.scopus.com/authid/detail.uri?origin=AuthorProfile{\&}authorId=6603935934{\&}zone=} shares a profile with others, and 
despite the fact\footnote{At the time of collecting the data for this article.} that he has written~$27$ articles gathering a total of~$219$ citations, his mixed 
profile mentions~$163$ articles and~$6129$ citations. Similar cases appear for AUEB\_DI's PhD George Tsatsaronis, recording~$286$ articles and~$5074$ citations 
instead of the `correct' $32$~articles and~$382$ citations, for CRETE\_CSD' Maria Markaki, and so on. However, there were many cases where the supervisor had no 
joint publications with some of his/her PhDs. We even encountered a case where the supervisor had no joint publications with any of his PhDs(!). In this case, we 
had no alternative but to use Scopus `Search Form', and deal with author disambiguation and mixed profiles issues.

The profile data used in this study, i.e., number of publications, number of citations and $h$-index of each PhD, were retrieved by the respective Scopus profiles
during the days June~24-25, 2017. As far as we can tell, there were no updates in the Scopus database during these days, so the information gathered for all PhDs 
concerns the same `database instance'. An exception to the collection dates is the UNIPI\_DI whose data were collected in July 3rd. During this week there was 
a Scopus update, thus this department is slightly benefitted. However, since it performed moderately in the rankings, there is no substantial argument against 
the validity of the data as a whole. We need to emphasize here, that the Scopus profiles were discovered and processed (to perform author disambiguation, to 
cleanse `mixed' profiles) in earlier time; during the aforementioned days, we just retrieved the data of interest. Despite the breadth (in areas) and depth 
(in years) of the Scopus database, we could not find all PhDs Scopus profiles. Table~\ref{tab-percentages-missing-phds} shows the (approximate) percentage of 
missed profiles in our study. This is due to one or a combination of the following reasons: a) some PhDs have no profile at all in Scopus, because they have 
published no article(!) in any forum, i.e., journal or conference or collection indexed by Scopus (this holds for both recent, after $2005$, and older, before 
$2000$ PhDs); b) the articles of some PhD are quite old (around $1995$) and the Scopus data are sparse for this period; c) the authors of the present article 
were not able to do proper browsing and/or form an appropriate query to find the profile -- this is due to the really very different ways used by PhDs to spell 
their names, for instance `John' versus `Ioannis' versus `Giannis' versus `Yannis' \footnote{But, not `Yanis' (with one `n').} , `Cutsuridis' versus 
`Koutsouridis', `Economou' versus `Oikonomou, `Konstantinopoulos' versus `Constantinopoulos', `Sigletos' versus `Sygletos, and so on. Nevertheless, 
the missed profiles comprise a very small part of the total profiles, and we are confident that missing them will not affect the obtained results.

\begin{table}[!hbt]
\center
\begin{tabular}{||l|c||}\hline\hline
{\bf Department} & {\bf \% missing PhDs Scopus profiles}\\\hline
AEGEAN\_ICSD     & $0$\%\\\hline
AUEB\_DI         & $3.90$\%\\\hline
AUTH\_DI         & $1.53$\%\\\hline
AUTH\_ECE        & $1.16$\%\\\hline
CRETE\_CSD       & $1.33$\%\\\hline
CRETE\_ECE       & $0$\%\\\hline
IOAN\_CSE        & $0$\%\\\hline
IONIO\_DI        & $0$\%\\\hline
NTUA\_ECE        & $4.15$\%\\\hline
PATRAS\_ECE      & $2.05$\%\\\hline
PELOP\_DI        & $0$\%\\\hline
UA\_DI           & $0$\%\\\hline
UNIPI\_DI        & $6.82$\%\\\hline
UTH\_DIB         & $0$\%\\\hline
UTH\_ECE         & $0$\%\\\hline\hline
\end{tabular}
\vspace*{.05\baselineskip}
\caption{Percentage of PhDs Scopus profiles not available per department.}
\label{tab-percentages-missing-phds}
\end{table}

\section{Related work}
\label{sec-related-work}

There exist several articles whose focus is the evaluation of Hellenic universities faculty's research quality using bibliometric data. After our pioneering 
work~\cite{Katsaros-PCI08}, many articles evaluated national departments of computer science/engineering~\cite{Altanopoulou-QHE12,Pitsolanti-arxiv17} (or 
international~\cite{Khalifa-WIS14,Garousi-CIS12}), national chemical engineering departments~\cite{Kazakis-Scientometrics15,Lazaridis-Scientometrics10}, national 
civil engineering departments~\cite{Kazakis-Scientometrics14b}, economics departments~\cite{Katranidis-BER14}, medical schools~\cite{Kazakis-Scientometrics14}, 
or national and international departments from various -- such as pedagogical, technological, political science, sociology, marketing -- 
disciplines~\cite{Altanopoulou-QHE12,Miroiu-QHE15,Vaxevanidis-MEE12,Vaxevanidis-IJQR11}.
 
To our knowledge this is the first article focusing in the evaluation of university departments' PhD research programmes under the prism of productivity and 
impact quantification of the research work conducted in the `scientific lifetime' of their former PhDs. The latent hypothesis of this article is that the research 
capacity of a PhD is (partly or significantly) shaped during his PhD studies.

\section{Conclusions}
\label{sec-conclusions}

This article has conducted a bibliometric evaluation of~$15$ departments of computer science/engineering of Hellenic universities. As a conclusion to the article
we would like to record only the three most significant findings of our study: a) there is no correlation among graduation time and impact of the work of a PhD, 
b) there is no evident\footnote{In a future study, this can be accurately quantified.} correlation among productivity and impact either in the supervisors or in 
the PhDs realm, and finally c) there is a clear `winner' department across all our measures, and that is CRETE\_ECE, followed by AUTH\_DI and CRETE\_CSD.

\subsection*{Acknowledgements}

The authors are indebted to the following people who offered invaluable help in providing access to raw data concerning PhD graduates of their department: 
Vassilios Dimakopoulos, Apostolos Dollas, Rania Doufexi, Vana Doufexi, Ioannis Fudos, Antigoni Galanaki, Vicky Grigoraki, Dimitris Gritzalis, George Hassapis, 
Yannis Ioannidis, Elma Kalogeraki, Mary Karasimou, Sofia Kataki, Nectarios Koziris, Elena Laskari, Kity Loupa, Grigoris Papagiannis, Antonis Paschalis, 
Vassilis Plagianakos, Spyros Sioutas, Spiros Skiadopoulos, Angeliki Spyrou, George~D. Stamoulis, George Tsihrintzis, Dimitrios Vlachos, Michalis Zervakis. 
The first author wishes to thank Nikos Bellas, his colleague at the Department of Electrical and Computer Engineering of the University of Thessaly, 
whose thought-provoking comments and observations motivated the present work.

\bibliographystyle{plain}
\bibliography{phdgr}

\vspace*{-6cm}
\begin{IEEEbiography}[{\vspace*{-1.5\baselineskip}\includegraphics[height=1.1in,clip,keepaspectratio]{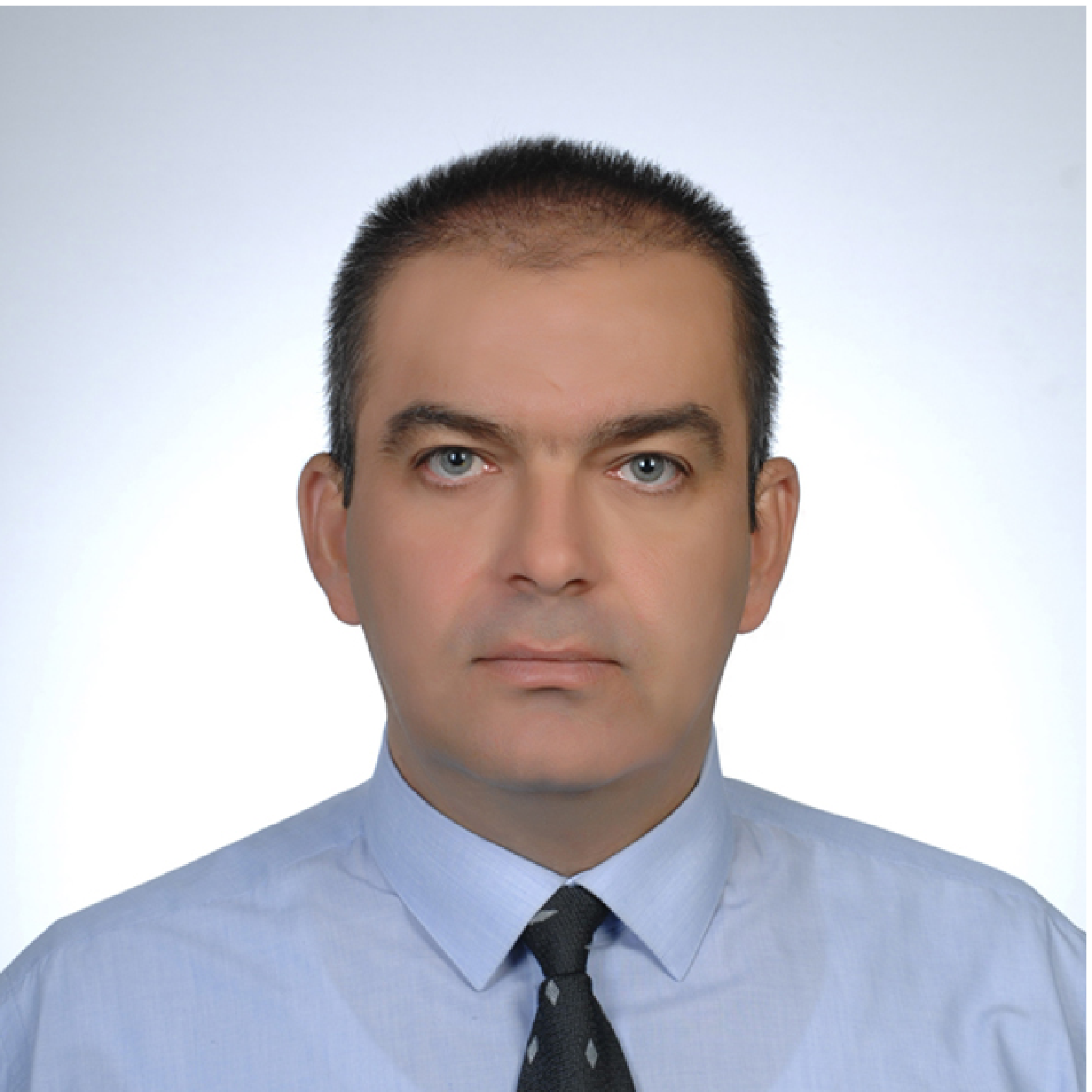}}]{Dimitrios Katsaros}
%is visiting assistant professor in the Department of Electrical Engineering at Yale University, and
is assistant professor in the Department of Electrical and Computer Engineering at the University of Thessaly. 
%His research interests include distributed systems.
Katsaros can be reached at dkatsar@e-ce.uth.gr or d.katsaros@yale.edu
\end{IEEEbiography}
\vspace*{-7cm}
\begin{IEEEbiography}[{\vspace*{-1.95\baselineskip}\includegraphics[height=1.2125in,clip,keepaspectratio]{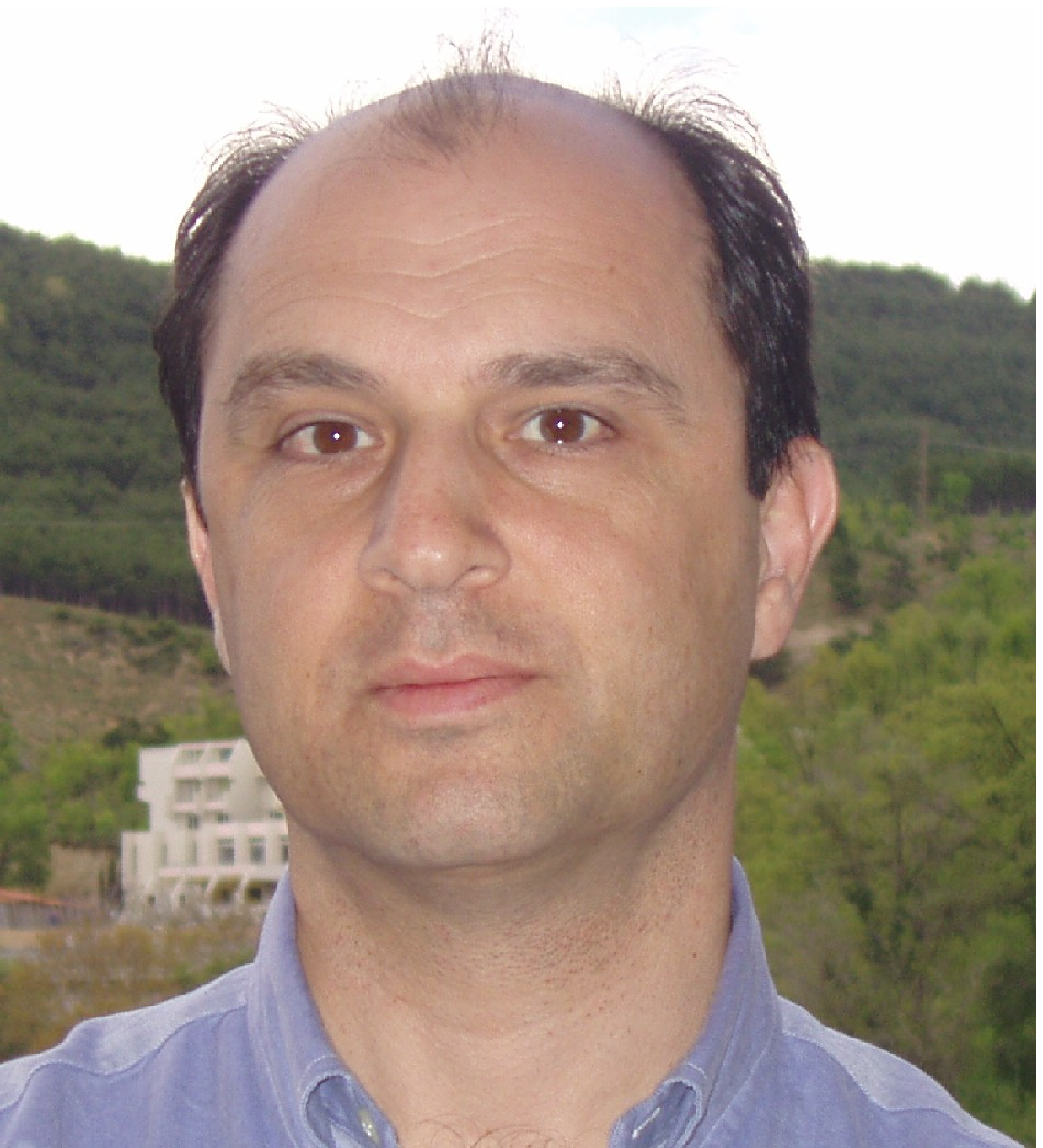}}]{Yannis Manolopoulos}
is professor in the Department of Informatics at Aristotle University of Thessaloniki.
%Manolopoulos has been with the University of Toronto, the University of Maryland at College Park and 
%the University of Cyprus. He has also served as Rector of the University of Western Macedonia, Greece, 
%Head of his own department and Vice-Chair of the Greek Computer Society. 
%His research interests include database systems, the Web, and data mining.
Manolopoulos can be reached at manolopo@csd.auth.gr
\end{IEEEbiography}

\end{document}